\documentclass[12pt,preprint]{aastex}         % preprint format, AASTEX 5.0
\usepackage{graphicx}
\usepackage{epstopdf}
\usepackage{amsmath}
\def\be{\begin{equation}}
\def\ee{\end{equation}}

\def\lambar{\lambda\llap {--}}

\def\tt{\mbox{\boldmath $\theta $}}
\def\pp{\mbox{\boldmath $\phi $}}

\def\lambar{\lambda\llap {--}}
\def\lsim{\lower 2pt \hbox{$\, \buildrel {\scriptstyle <}\over
         {\scriptstyle \sim}\,$}}

\begin{document}
\newcommand{\figureout}[3]{\psfig{figure=#1,width=5.5in,angle=#2} 
   \figcaption{#3} }

\title{Synchrotron Self-Compton Emission from the Crab and Other Pulsars}

\author{Alice K. Harding\altaffilmark{1} and Constantinos Kalapotharakos\altaffilmark{1}$^{,}$\altaffilmark{2}} 
  
\altaffiltext{1}{Astrophysics Science Division,      
NASA Goddard Space Flight Center, Greenbelt, MD 20771}
\altaffiltext{2}{University of Maryland, College Park (UMDCP/CRESST), College Park, MD 20742}
 
%\slugcomment{To appear in The Astrophysical Journal, *** issue.}

\begin{abstract}
Results of a simulation of synchrotron-self Compton (SSC) emission from a rotation-powered pulsar are presented.  The radiating particles are assumed to be both  accelerated primary electrons and a spectrum of electron-positron pairs produced in cascades near the polar cap.  They follow trajectories in a slot gap using 3D force-free magnetic field geometry, gaining pitch angles through resonant cyclotron absorption of radio photons, radiating and scattering synchrotron emission at high altitudes out to and beyond the light cylinder.  Full angular dependence of the synchrotron photon density is simulated in the scattering and all processes are treated in the inertial observer frame.  Spectra for the Crab and Vela pulsars as well as two energetic millisecond pulsars, B1821-24 and B1937+21 are simulated using this model.   The simulation of the Crab pulsar radiation can reproduce both the flux level and the shape of the observed optical to hard X-ray emission assuming a pair multiplicity of $M_+ = 3 \times 10^5$, as well as the very-high-energy emission above 50 GeV detected by MAGIC and VERITAS, with both the synchrotron and SSC components reflecting the shape of the pair spectrum.  Simulations of Vela, B1821$-$24 and B1937+21, for $M_+$ up to $10^5$, do not produce pair SSC emission that is detectable by current telescopes, indicating that only Crab-like pulsars produce significant SSC components.  The pair synchrotron emission matches the observed X-ray spectrum of the millisecond pulsars and the predicted peak of this emission at 1 - 10 MeV would be detectable with planned Compton telescopes.

\end{abstract} 

%\keywords{theory --- pulsars: general --- stars: neutron}

\pagebreak
%\received{}
%\revised{}      
%\accepted{}
  
\section{Introduction}

The recent detection of pulsed emission from the Crab and Vela pulsars at energies above 50 GeV by ground-based air-Cherenkov telescopes 
marks a turning point both in technological achievement of the detectors and in pulsar acceleration and radiation modeling.  The MAGIC telescope
first detected pulsed emission from the Crab above 25 GeV (Aliu et al. 2008), followed by pulsed detections above 100 GeV by VERITAS (Aliu et al. 2011) and up to 400 GeV with MAGIC (Aleksic et al. 2012).   The averaged and phase-resolved spectra above 25 GeV are consistent with a smooth broken power-law extension of the spectrum measured by the {\it Fermi} Large Area Telescope (LAT) above 100 MeV, thus ruling out an exponentially cutoff power-law spectrum predicted by curvature radiation (CR) models for GeV emission (e.g. Romani 1996, Harding et al. 2008 [HSDF08]).  Very recently, pulsed emission was detected from the Vela pulsar above 30 GeV with the HESS (Stegmann et al. 2014) telescope and above 50 GeV with the {\it Fermi}-LAT (Leung et al. 2014).  In this case it is not clear whether another emission component in addition to CR is required.  Most recently, VERITAS has reported a non-detection of the Geminga pulsar above 100 GeV (Aliu et al. 2015).   McCann (2015) performed a stacking analysis of 115 {\it Fermi} pulsars from the {\it Fermi}-LAT 2nd Pulsar Catalog (Abdo et al. 2013), both young and MSPs, and found no significant emission at more than 7\% Crab level  above 50 GeV.

For many years, CR was the favored mechanism for the high energy ($> 100$ MeV) pulsed emission in a number of different models for pulsar emission, and VHE emission similar to what has been detected from the Crab was never predicted.   In models where acceleration and emission occur above the pulsar polar caps (PC models, Daugherty \& Harding 1996), the predicted spectrum of CR from primary electrons has a very sharp ``super-exponential" cutoff due to attenuation by the magnetic pair production process that occurs in the very strong magnetic fields near the neutron star surface.  Measurement of the spectral cutoff of the Vela pulsar by {\it Fermi} (Abdo et al. 2009) has shown that a super-exponential cutoff can be eliminated with high significance, ruling out emission from near the neutron star PCs.  Models for emission originating in the outer magnetosphere are therefore favored, since many of these predict 
curvature radiation spectra from primary particles with exponential spectral cutoffs at energies of a few GeV that agree well with 
those measured for most pulsars by {\it Fermi}.  Such models include the outer gap (OG), where acceleration and emission occur in vacuum gaps that form above the null charge surface (e.g. Romani 1966, Cheng et al. 2000, Hirotani 2006) and the slot gap (SG), where acceleration and emission occur in a narrow region along the last open field lines from the neutron star surface to near the light cylinder (Muslimov \& Harding 2004 [MH04], HSDF08).  More recently, models postulating high-energy emission originating outside the light cylinder, near or in the current sheet that forms in the spin equatorial plane have been developed.  Several of these models assume that synchrotron radiation (SR) from particles accelerated by magnetic reconnection in the current sheet produces the GeV emission (e.g. P\'{e}tri 2012, Uzdensky \& Spitkovsky 2014), while others assume that CR by particles accelerated by induced electric fields in dissipative magnetospheres produces the GeV emission (Kalapotharakos et al. 2014, Brambilla et al. 2015).  

The {\it Fermi} LAT has to date detected more than 160 $\gamma$-ray pulsars and all of these with statistics adequate for spectral fitting of a power law with exponential cutoff
\be
dN(E)/dE = E^{-\gamma}\,\exp[-(E/E_c)^b],
\ee
show cutoffs, $E_c$, in the range 1 - 7 GeV (Abdo et al. 2013), assuming $b = 1$.  For a subset of pulsars having higher counts, fits to the above expression with $b$ free can be performed and a number of these are best fit with $b < 1$, including the Crab and Vela pulsars, indicating a more gradual cutoff than pure exponential.  This finding raises the question of whether the more gradual high-energy cutoffs are due to a blending of CR spectra from a number of different rotation phases (Abdo et al. 2010) and/or locations in the magnetosphere (Leung et al 2014), or additional emission components.  It has been suggested that the very-high-energy (VHE) emission from the Crab is inverse-Compton emission (Lyutikov et al. 2012), since it is difficult to produce photons above 100 GeV with CR.  Lyutikov (2013) modeled the Crab X-ray and $\gamma$-ray spectra as cyclotron-self Compton emission from electron-positron pairs in the outer magnetosphere, fitting the data to determine the spectrum and multiplicity of the pair spectrum as well as the location of emission.  An synchrotron-self Compton (SSC) model for the Crab VHE emission from the outer gap (OG) was suggested by Aleksic et al. (2011) and Du et al. (2012) modeled the observed Crab emission from the annular gap.  Alternatively, it was proposed that particles accelerated by reconnection in the current sheet reach temperatures of 10 GeV, if equipartition is assumed, and their SR can extend to 100 GeV in the observer frame after Doppler boosting (Uzdensky \& Spitkosky 2014, Iwona \& Petri 2015).   

In this paper, we apply a pair cascade simulation coupled to an outer magnetosphere radiation model to predict the broadband X-ray to VHE $\gamma$-ray spectra of several classes of pulsar.   In this model, the pairs are produced in cascades above the PCs initiated by primary accelerated electrons and sustained by magnetic pair creation (e.g. Daugherty \& Harding 1982).  The pairs lose their momentum perpendicular to the magnetic field to SR near the PCs and then stream into the outer magnetosphere, where they  absorb radio photons in the cyclotron resonance to regain pitch angles and emit SR (HSDF08).  The SR provides soft photons for inverse-Compton scattering by these same particles to boost their energies to the $\gamma$-ray range.  We also include CR radiation of primaries accelerated by electric fields in a SG geometry, as well as their SR components due to cyclotron resonant absorption.  The radio emission is simulated using an empirical cone and core beam model.  Since the magnetic field structure in the outer magnetosphere becomes important for these models, we use a force-free magnetosphere configuration to define the magnetic field, a significant improvement over the retarded vacuum dipole field that we had used previously (HSDF08).  Use of the force-free magnetic field structure allows us to follow particles and radiation arbitrarily close to and even beyond the light cylinder, which should not be a physical limitation of emission models.  Although force-free magnetospheres assume that the electric field parallel to the magnetic field, $E_\parallel$, vanishes, they well approximate the field structure of energetic pulsars where departures from force-free conditions exist only in narrow accelerator gaps.  We therefore use the force-free field structure but assume acceleration of primary particles in a narrow SG.  Using these simulations, we can address the question of how many pulsars should have detectable SSC emission and how high in energy this emission extends.

\section{3D Magnetosphere Geometry} \label{sec:mag}

Since radiation of relativistic particles is beamed along their direction of momentum, and particle trajectories in pulsar magnetospheres follow closely (but not exactly) the magnetic field lines, the structure of the magnetic field is a critical component to emission models.  A static dipole is not a good approximation to a pulsar magnetic field, even near the neutron star, since the rotation causes a sweepback of the field lines near and beyond the light cylinder.  The vacuum retarded dipole solution (Deutsch 1955), adopted in many radiation models (e.g. Romani \& Yadigaroglu 1995, Cheng et al. 2000, HSDF08), incorporates the sweepback due to retardation of the field and produces a distortion and shift of the open field volume (including the PC) toward its trailing edge (Dyks \& Harding 2004).  Numerical solutions of force-free magnetospheres (Spitkovsky 2006, Timokhin 2006), where the plasma density is assumed to be high enough to screen $E_\parallel$ so that the ideal MHD condition ${\bf E \cdot B = 0}$ is enforced, show a larger degree of magnetic field sweepback, as well as straightening of the poloidal field lines due to currents.  Recent simulations of dissipative pulsar magnetospheres (Kalapotharakos et al. 2012, Li et al. 2012), that drop the ideal MHD condition and allow finite $E_\parallel$, show that the structure of the magnetic field stays close to force-free for the high plasma conductivity needed to match the $\gamma$-ray light curves (Kalapotharakos et al. 2014, hereafter KHK14) and spectra (Brambilla et al. 2015) of young pulsars.  

We will therefore adopt the force-free magnetosphere structure to model the particle trajectories and radiation, and all calculations are done in the non-rotating, inertial observer's frame (IOF).  The pairs are assumed to be produced at the PCs and to screen the $E_\parallel$ at higher altitudes, except for a narrow gap along the last open field lines (the SG).  They will therefore not experience any acceleration in the calculation of their radiation in the global magnetosphere.  The primary electrons are assumed to be accelerated only in the SG, with a constant $E_\parallel$ that we take as a free parameter.  Both pairs and primary electrons are injected at the neutron star surface at footprints of open field lines that are determined by ``open volume coordinates" (Dyks et al. 2004 [DHR04]).  These coordinates map the open field region bounded by field lines that close within the light cylinder, $R_{\rm LC} = c/\Omega$, where $\Omega$ is the pulsar angular rotation frequency.   The magnetic field vector is determined at each point by 3D interpolation of a table of values read-in from a numerical force-free magnetosphere solution (Kalapotharakos et al. 2012).   This solution has a resolution of $0.02\,R_{\rm LC}$ and extends to a minimum radius of $0.2\,R_{\rm LC}$, which is the surface of a pulsar with period $0.8$ ms.  To extend this solution to the surface of pulsars with longer periods, we join the numerical solution to a retarded vacuum dipole solution over the range $0.2 - 0.4 \,R_{\rm LC}$ using a ramp function.  

The code first computes the PC rim at the neutron star surface by performing 4th order Runge-Kutta integrations along the field lines of the numerical solution to determine the open field footpoints.  This is done by iteration, choosing an initial value of magnetic polar angle $\theta_i$ at a number of different values of magnetic azimuth $\phi$.  If the field line with that footpoint does/does not close within $R_{\rm LC}$, a smaller/larger value of $\theta$ is chosen until the closure point is at $R_{\rm LC}$ within a given tolerance.  We define ``open volume coordinates" $(r_{\rm ovc}, l_{\rm ovc})$ to identify footpoints at the neutron star surface of field lines along which the particles are traced.  The ``radial" coordinate $r_{\rm ovc}$, equivalent to magnetic polar angle,  is equal to 0 at the magnetic pole and 1 at the PC rim.  Lines of constant $r_{\rm ovc}$ define a set of concentric deformed rings that fill the PC surface (see Figure 2 of DHR04).  The ``azimuthal" coordinate $l_{\rm ovc}$ measures the arclength along each distorted ring with fixed $r_{\rm ovc}$.  The $l_{\rm ovc}$ increase in the direction of magnetic azimuth, $\phi_{pc}$,  in a counterclockwise direction around the PC, starting from $l_{\rm ovc} = 0$ at $\phi_{pc} = 0$, defined to be at the magnetic meridian (i.e. the line between the magnetic and spin axes).  With this definition, $\phi_{pc} = 3\pi/2$ is on the leading side of the PC and $\phi_{pc} = \pi/2$ is on the trailing side (Note that $\phi_{pc} = \phi + \pi$).

To model an injection of particles that is uniform over the PC, we use a set of rings between $r_{\rm ovc}^{\rm min}$ and $r_{\rm ovc}^{\rm max}$ with equal spacing $d_{\rm ovc}$.  These footpoints are then spaced uniformly in each of the rings with $N_l = \Delta_{\rm azim}360$ equal divisions.  Since each ring has the same number of footpoints, and the rings have varying circumferences, the contribution from different rings must be weighted by $l_{\rm ring}/l_{\rm rim}$,where $l_{\rm ring}$ is the length of a particular ring and $l_{\rm rim}$ is the total PC rim length.  To determine the trajectories of particles in the IOF, we require the total particle velocity to be the sum of a drift component and a component parallel to the magnetic field (KHK14),
\be
\label{velocityv}
\mathbf{v}=\left(\frac{\mathbf{E}\times\mathbf{B}}{B^2+E_0^2}+f
\frac{\mathbf{B}}{B}\right)c~,
\ee
where (Gruzinov 2008, Li et al. 2012),
\begin{eqnarray}
E_0^2 = B_0^2 - \mathbf{B}^2 + E^2 \\ \nonumber
\\
B_0^2 =( \mathbf{B}^2 - \mathbf{E}^2 + \sqrt{(\mathbf{B}^2 - \mathbf{E}^2)^2 + 4(\mathbf{B} \cdot \mathbf{E})^2} )/2 \\ \nonumber
\\
B_0 = {\rm sign}(\mathbf{B} \cdot \mathbf{E})\sqrt{B_0^2}.
\end{eqnarray}
By requiring that ${\bf v} \simeq c$ and that the motion is outward, we can solve for the scalar quantity $f$ at each point along the trajectory.  In the special case of a force-free magnetosphere, ${\bf v}$ stays parallel to $B$ in the corotating frame.  In the IOF, the step size along the particle trajectory is $d\ell = ds/f$, where $ds$ is the 
step size along the rotating magnetic field line (see Appendix C of Bai \& Spitkovsky 2010).  In treating the particle trajectories in the IOF with force-free field geometry, it is possible to extend the trajectories and radiation outside the light cylinder.  Although the particles are following the field, they slide forward along the field lines as the field lines sweep back, following a nearly radial path in the IOF at $r > R_{\rm LC}$.

\section{Pair Cascade Spectra} \label{sec:pair}

The process of electron-positron pair production in pulsar magnetospheres is thought to be critical for generating charges for the magnetosphere and for the pulsar wind nebula, as well as plasma necessary for coherent radio emission.  The pairs are produced in electromagnetic cascades above the PCs (Daugherty \& Harding 1982) or in OGs (Cheng et al. 1986).   Here, we assume that the pairs that radiate SR and SSC in the outer magnetosphere originate from PC cascades, that are initiated in the strong electric fields near the neutron star surface by acceleration primary electrons.  We have calculated the spectra of pairs using a code that simulates a steady (non-time-dependent) electromagnetic cascade above the pulsar PC (Harding \& Muslimov 2011 [HM11]).  Primary particles are accelerated by an electric field induced by rotation of the magnetic field, derived assuming space-charge-limited flow (i.e., free emission of particles from the neutron star surface) (Arons \& Scharlemann 1979), and emit curvature radiation.  The highest energy curvature photons are absorbed by magnetic pair attenuation (Erber 1966, Daugherty \& Harding 1983), producing a spectrum of first-generation electron-positron pairs.  The pairs are born in excited Landau states (with non-zero pitch angles) and radiate synchrotron photons that spawn further generations of pairs. Since the total cascade multiplicity $M_+$ (average number of pairs produced by each primary particle) depends on the pulsar period $P$ and surface magnetic field strength $B_{\rm s}$, younger pulsars produce high $M_+$ cascades but older pulsars, with low magnetic fields and long periods, have much lower $M_+$.   This results in a pair death line in $P$-$\dot P$ which, in the case of a dipole field, predicts that the older half of the pulsar population cannot produce significant pair multiplicity.

However, the sweepback of magnetic field lines near the light cylinder (due to real and displacement currents), as well as possible field distortions near the neutron star will cause the magnetic PCs to be offset from the dipole axis (e.g. Dyks \& Harding 2004). HM11 introduced a perturbed dipole field structure with non-axisymmetry and showed that the resulting offset PC enhances pair multiplicity by increasing the accelerating electric field.   For the pair cascade simulations, we will adopt this magnetic field structure which has two configurations for the dipole offset in which the magnetic field is symmetric or asymmetric with respect to the dipole axis.  Modeling of the thermal X-ray light curves of several MSPs requires an offset of the PC hotspot that is symmetric (Bogdanov et al 2013).  We adopt the symmetric distorted field in simulating the pair spectra of the MSPs in this paper.  In this case, the magnetic field in spherical polar coordinates ($\eta $, $\theta $, $\phi $) is
\noindent 
\begin{equation}
{\bf B} \approx {B_{\rm s}\over {\eta ^3}}~\left[ \hat{\bf r}~\cos \theta +{1\over 2} ~\hat{\tt}~ (1+a)~\sin \theta - \hat{\pp }~\epsilon ~\sin \theta~\cos \theta~\sin (\phi - \phi_0) \right],
\label{B1}
\end{equation}
where $\eta = r/R$ is a dimensionless radial coordinate in units of the stellar radius $R$, $B_{\rm s}$ is the surface field strength at the magnetic pole, $a=\epsilon ~\cos (\phi - \phi _0)$ is a parameter defining the azimuthal distortion of the polar field lines from a dipole, and $\phi_0$ is the magnetic azimuthal angle that defines the plane of the PC offset.  HM11 derive the parallel component of the electric field, $E_\parallel$, using this field structure and we have used the $E_\parallel$ of Eqn (11) of HM11 that corresponds to a symmetric offset.
We use this $E_\parallel$ to accelerate the electrons  from different starting points over the PC surface and simulate the pair cascades over the whole PCfor a range of $P$, $\dot{P}$ (or equivalently, $B_{\rm s}$), and offset parameter $\epsilon$.  We assume a non-zero offset only for the MSPs.  Pair spectra for the parameters of the Crab ($\epsilon = 0$) , Vela ($\epsilon = 0$) and MSPs PSR B1821-24 ($\epsilon = 0.6$, $\phi_0 = 3\pi/2$) PSR B1937+21 ($\epsilon = 0.6$, $\phi_0 = 3\pi/2$) are shown in Figure \ref{fig:pairspec}.  The Crab pair spectrum extends over five decades of pair energy, from $\gamma_+ = 20$ to $\gamma_+ = 10^6$, while the Vela pair spectrum has a smaller range, cutting off around $\gamma_+ = 4 \times 10^5$.  The difference is due to the shorter period and larger PC size of the Crab, which gives smaller radii of curvature of the field lines,  so that CR photons have higher energy and gain large angles to the field,  and the higher magnetic field strength.  The pair spectra of the MSPs start at energies about two decade higher, $\gamma_+ \sim 2 \times 10^3$ and extend to higher energy, $\gamma_+ \sim 10^7$.  This striking difference in pair spectra of young and MSPs comes from their large difference (four orders of magnitude) in surface field strengths.  As described by HM11, the photons must have higher energies to produce pairs in lower magnetic fields, which raises the minimum pair energy,  but are also able to escape at higher energies, raising the high energy cutoff in the pair spectrum.  The non-zero PC offset decreases the lower cutoff of the pair spectrum relative to the pair spectrum with zero offset by a fraction of a decade.

Simulations of PC cascades including time dependence of the electromagnetic fields and plasma production (Timokhin 2010, Timokhin \& Arons 2013) find that to satisfy the charge and current density of the global magnetosphere models the pair cascades are non-steady.  The accelerating electric field is time-dependent, in the general case where the magnetospheric current density $J$ is not very close to the Goldreich-Julian value, $J_{GJ} = \rho_{GJ} c$, with cycles of pair creation followed by complete screening of the electric field.  The timescale of these pair cascade cycles is very short (of order the light crossing time across the gap).  Such non-steady pair cascades can produce pair multiplicities that are much higher than the steady cascades.  For young pulsars such as the Crab and Vela, the time-averaged pair multiplicities are about several times $10^5$  (Timokhin \& Harding 2015), compared to $\sim 2 \times 10^4$ for the Crab and $\sim 6 \times 10^3$ for Vela steady pair cascades.  Time-dependent pair cascade results are not yet available for MSPs, where 2D simulations are required.  Although it is beyond the scope of this paper to simulate time-dependent pair spectra (especially for millisecond pulsars [MSPs]), we will allow for pair multiplicities higher than that of steady cascades in our SSC simulations.

\section{Simulation of Emission}

We model the high-energy radiation over the entire spectrum from optical to VHE $\gamma$-ray wavelengths.  Emission from both primary particles and pairs are simulated as they move along their trajectories with step size $\Delta \ell$, starting at the neutron star surface to a maximum radius $r_{\rm max}$.  Primary electrons are assumed to undergo continuous acceleration with a constant electric field, $d\gamma/d\ell = R_{\rm acc}$ in a narrow gap along the boundary of the open field region, between $r_{ovc} = 0.95 - 0.99$.   The electric field in the high-altitude gap is not the same field from HM11 that is used for the pair cascades, since the HM11 field is only valid for low altitudes near the PC.  The high-altitude gap forms at the outer edge of the PC, where the interior $E_\parallel$ decreases to values too low for pair cascades to develop.  Although analytic solutions exist for both the low altitude SG (Muslimov \& Harding 2003) and its extension to high altitudes (MH04), we have chosen not the use the solution of MH04 for several reasons.  First, the gamma-ray luminosity modeled using the SG $E_\parallel$ solution of MH04  with widths narrow enough to produce the observed light curves falls short of the observed luminosities by about a factor of 10 for young pulsars.  As noted by Pierbattista et al. (2012), the $E_\parallel$ solution for the original SG and also the OG do not provide enough gamma-ray luminosity to match the observed {\it Fermi} pulsars.  The $E_\parallel$ of HM2011 on the offset side of the PC is larger and may produce a large enough $E_\parallel$ at high altitude to produce luminosities matching the data.  However, it is very difficult to derive the $E_\parallel$ extension for the offset PCs and this has not yet been done.  Also, with the advent of dissipative magnetosphere models we will soon be able to have more self-consistent $E_\parallel$ throughout the magnetosphere, albeit numerical.  Until these are available, we decided it is better to treat $E_\parallel$ at high altitudes as a free parameter.
We have therefore simply assumed a constant $E_\parallel$ for the high-altitude gap, similar to the behavior of the high-altitude symmetric SG MH04.

\begin{figure} 
\includegraphics[width=180mm]{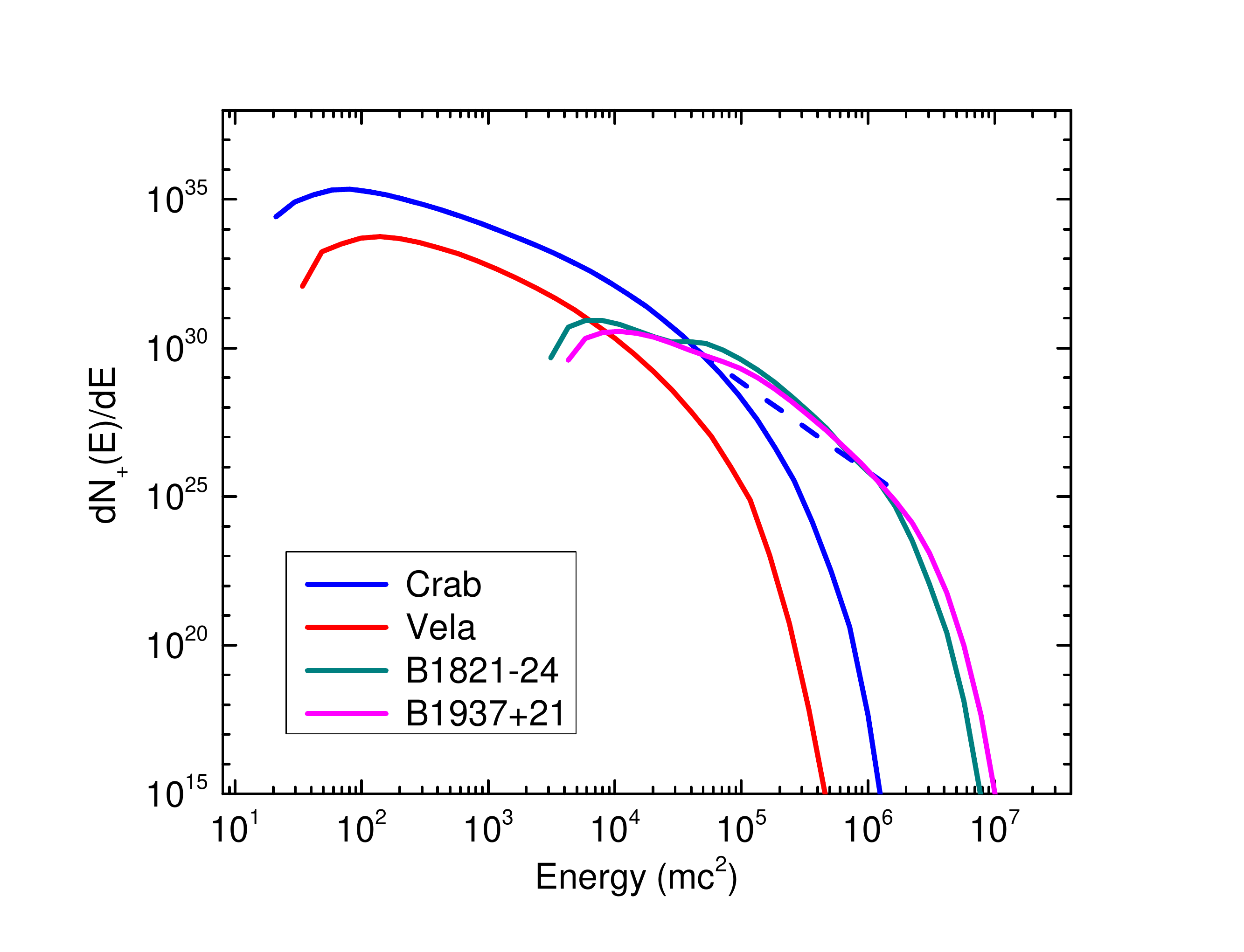}
\caption{Pair spectra (in units of  pairs/s/mc$^2$) from polar cap pair cascade simulations for the Crab, Vela, B1821-24 and B1937+21 pulsars, as labeled.  The dashed line is a power law extension to the Crab pair spectrum (see text).}   %fig. 1
\label{fig:pairspec}
\end{figure}

The electron-positron pairs are assumed to experience no acceleration as they flow along their trajectories in the screened region inside the gap, between $r_{ovc} = 0.91 - 0.95$.  CR, SR and SSC radiation are simulated for all particles in the same way, as described below.  The primary electrons are injected at the neutron star surface with very low Lorentz factors, $\gamma = 2$, while the pairs are injected with the pair cascade spectra shown in Figure \ref{fig:pairspec}.  All particles are assumed to initially have momentum only parallel to the magnetic field, since the pairs lose all their perpendicular momentum to SR very near the neutron star surface in the pair cascade.  
Single primary electrons are injected at the start of each primary trajectory at the neutron star surface and their radiation is normalized to the primary flux 
\be \label{eq:np}
\dot n_p = n_{\rm GJ}\, c\, \pi R^2\,\theta_{PC}^2\,[(r_{ovc}^{\rm max})^2 - (r_{ovc}^{\rm min})^2]
\ee
where $n_{\rm GJ} = B_0\Omega/2\pi e c$ is the Goldreich-Julian density and $\theta_{PC} = (\Omega R/c)^{1/2}$ is the PC half-angle.  
 The primary particles from the interior of the PC are on field lines that do not thread the high-altitude accelerator gap, but are on field lines where the accelerating field is screened above a pair formation front by pair cascades at low altitude.  Since these primaries are not accelerated above the pair front, they lose most of their energy quickly and do not contribute much to emission at high altitudes.  The emission from the PC pair cascades is all emitted at low altitudes, below 6-7 stellar radii, and will only be visible when an observer viewing angle is very close to the magnetic axis.  Since most observer angles will not see the PC emission, we therefore can neglect their contribution in the present study.

To simulate radiation from a spectrum of pair energies, shown in Figure \ref{fig:pairspec}, the pair spectrum is divided into logarithmic energy intervals, $\Delta \gamma_+$, between $\gamma_+^{min}$ and $\gamma_+^{max}$ and a single pair member (electron or positron) at each energy, $\gamma_+$, is injected at the base of each pair trajectory.  Their radiation is normalized to the flux of pairs in each energy interval,
\be \label{eq:n+}
\dot n_+ (\gamma_+) = {2 M_+ n_+ (\gamma_+) \Delta\gamma_+ \dot n_p \over \int_{\gamma_+^{\sc min}}^{\gamma_+^{max}}\,n_+ (\gamma_+) d\gamma_+}
\ee
and $M_+$ is the pair multiplicity.  We have kept $M_+$ as an independent quantity rather than derive it from the integral of the pair spectrum so that it can be treated as a free parameter in the simulation.

The direction of photon emission, $\boldsymbol{\eta_{em}}$, is along $\boldsymbol{\beta} = {\bf v}/c$, the direction of particle motion in the IOF, with photon emission angles 
\be
\mu_{em} = \beta_{z}, \,\,\, \phi_{em} = {\rm atan}\left({\beta_{y}\over \beta_ {x}}\right).
\ee
As the particle moves a distance $d\ell$, the change in phase is $\Delta\phi_{\rm rot} = \Omega d\ell/c = \Omega dt$.  The emitted photons are accumulated in sky maps of observer angle $\zeta_{\rm obs} = {\rm acos}(\mu_{em})$ versus phase $\phi_{\rm obs}$, both with respect to the pulsar spin axis (the $z$ axis on the IOF Cartesian grid).  To determine the observed phase, we apply the rotation and add the time delay correction,
\be
\phi_{\rm obs} = \phi_{em} - \Delta\phi_{\rm rot} + {{\mathbf{r_{em}} \cdot  \boldsymbol{\eta_{em}}} \over R_{\rm LC}}
\ee
where $\mathbf{r_{em}}$ is the radius of photon emission.
The quantities in the sky maps are the emitted photon fluxes, normalized using Eqn (\ref{eq:np}) for primaries and Eqn (\ref{eq:n+}) for pairs, divided by the solid angle of that sky map bin, $\Delta\Omega = \sin\zeta_{\rm obs}\Delta\zeta_{\rm obs}\Delta\phi_{\rm obs}$.  The phase-averaged observed flux at a viewing angle $\zeta_{\rm obs}$ is then obtained by summing the fluxes in $\phi_{\rm obs}$ and dividing by $2\pi d^2$, where $d$ is the distance to the source.  

\subsection{Curvature Radiation} 

The energy spectrum of curvature radiation from a single electron with Lorentz
factor $\gamma$ is
\be
N_{CR}(\varepsilon) = \sqrt{3}\,{e^2\over c}\,\gamma\,\kappa\left({\varepsilon\over 
\varepsilon_{cr}}\right)
\ee
where $\varepsilon$ is the emitted photon energy in units of $mc^2$ and
\be
\varepsilon_{cr} = {3\over 2} {c\over \rho_c}\,\gamma^3,
\ee
and the function $\kappa(x)$ is defined as
\be
\kappa(x) \equiv x \int_{x}^{\infty} K_{5/3}(x')dx'.
\ee
The radius of curvature $\rho_c$ in the force-free magnetosphere is not that for a pure dipole field, but is the radius of curvature of the particle trajectory
determined in the IOF by computing the inverse of the trajectory curvature using the second derivative at the particle position:
\be
\rho_c = \left|{d^2\mathbf{v}\over d\ell^2}\right|^{-1}.
\ee

The approximate form of the photon spectrum is a power law with an exponential 
cutoff at $\varepsilon_{cr}$,
\be
{\dot N_{CR}(\varepsilon)\over d\varepsilon} = {\alpha \over (\lambar mc)^{1/3}} \left({c\over \rho_c}\right)^{2/3}\,
\varepsilon^{-2/3}\exp(-\varepsilon/\varepsilon_{\rm cr}).
\label{insCR} 
\ee

\subsection{Synchrotron Radiation} \label{sec:SR}

Since all particles start their trajectories with momentum parallel to the magnetic field, they will not radiate any SR until they acquire finite pitch angles.  We assume that particles gain pitch angles by resonant absorption of radio photons (Shklovsky 1970, Lyubarski \& Petrova 1998, LP98).  In this process, relativistic particles absorb photons that have energies at the cyclotron resonant frequency in their rest frame, which results in an excitation of the particle to a higher Landau state and an increase in pitch angle.  The particle will then emit spontaneous cyclotron radiation if its momentum perpendicular to the magnetic field is non-relativistic in the frame where the parallel momentum vanishes, or synchrotron radiation if the perpendicular momentum is relativistic.
The resonant absorption condition is 
\be  \label{rescond}
B' = \gamma\varepsilon_0\,(1-\beta\mu_0)
\ee
where $\gamma$ is the particle Lorentz factor, $\varepsilon_0$ is the lab frame energy (in units of $mc^2$) of the radio photon, 
$\beta = (1-1/\gamma^2)^{1/2}$, $B' = B/B_{\rm cr}$ is the local magnetic field in
units of the critical field strength $B_{\rm cr} = 4.4 \times 10^{13}$ G, 
$\mu_0 =\cos \theta_0$, and $\theta_0$  is the angle between the photon direction 
and the particle momentum in the lab frame . 
In pulsar magnetospheres, particles having large Lorentz factors will see radio photons at the cyclotron resonance at high altitude above the neutron star surface, when the local magnetic field has dropped low enough to satisfy the resonant condition,
(from equation [\ref{rescond}]),
\be
\gamma _R = 2.8\times 10^5 {{B_8}\over {\varepsilon _{0, {\rm GHz}} (1-\beta\mu_0)}}. 
\label{eq:gamma_R}
\ee
where $\mu_0$ is the incident absorption angle.

The particle initially undergoes absorption in low Landau states, where the
cyclotron emission rate is well below the absorption rate.  Therefore, the particle Landau state (and pitch angle) will increase 
stochastically but continuously until the increase in pitch angle through
resonant absorption and the decrease in pitch angle by synchrotron emission reaches an equilibrium.  This balance is not reached until the particle occupies a high Landau state, where it will radiate synchrotron emission.

We use the method of HSDF08, based on the work of LP98, to simulate the resonant cyclotron absorption and subsequent synchrotron emission from both primary electrons and pairs.  LP98 identified two regimes of particle pitch angle
increase in the resonant absorption.  In the first regime, when $\psi \ll \theta_0$ (the particle pitch angle, $\psi$, is less than the radio photon incident angle of the, $\theta_0$), the pitch angle is increasing while the momentum is nearly constant.  In the second regime, when  
$(\theta_0 - \psi) \ll \theta_0$, the pitch angle is constant as the total momentum increases.
In the regime where $\psi \ll \theta_0$ (equation (2.17) of Petrova (2002)), the mean square of the pitch angle increases as,
\be
\langle {\psi ^2} \rangle = 4\,\int _{\eta _R}^{\eta } {\it a_0}(\eta ')d\eta ' ,
\label{psi-msq}
\ee
where $\eta = r/R$, $\eta_{_R}$ is the radio emission altitude and
\be
{\it a_0}(\eta ) = {{2\pi ^2 {\it e}^2 (1-\beta\mu_0) I_0}\over {\gamma ^2 m^2c^4}} 
\left( {{\varepsilon _0 \gamma (1-\beta\mu_0)}\over {B'}} \right) ^{\nu },~~~~  \eta > \eta_{_R}.
\label{a0}
\ee 
Here $I_0$ is the observed intensity of the radio emission, in $erg\cdot cm^{-2}\cdot s^{-1}\cdot Hz^{-1}$
and $\nu$ is its spectral index. 
Thus, the change in perpendicular momentum due to cyclotron resonant absorption is
\be
\left({{dp_{\perp }}\over {dt}}\right)^{abs} = 2~{\it a_0(\eta)}~c {{\gamma ^2}\over {p_{\perp}}} + 
{{p_{\perp}}\over p} \left({{dp}\over {dt}}\right)^{abs} 
\label{dp_dt_eval}
\ee
where we used the assumption that $p_{\perp }$ is proportional to the root mean-square of the pitch angle,
$p_{\perp } = p {\langle {\psi ^2} \rangle}^{1/2}$.
We also compute the evolution of $p {\langle {\psi ^2} \rangle}^{1/2}$ 
rather than computing the evolution of the full particle distribution function.  

Combining Eqn (\ref{a0}) and (\ref{dp_dt_eval}), the resonant absorption rate is 
\be
\left({{dp_{\perp}}\over {dt}}\right)^{abs} = D{{\gamma^{\nu}}\over {p_{\perp }}} + 
{{p_{\perp } \gamma }\over {\gamma^2 -1}} \left({{d\gamma }\over {dt}}\right)^{abs},~~~~~~~~~\gamma < \gamma _R
\label{eq:dp_perp_abs}
\ee
where 
\be  \label{eq:D}
D = 5.7 \times 10^{9}\,\rm s^{-1}\,\gamma_R^{-\nu}\,\left({d_{kpc}\over \eta}\right)^2\,
\Phi_0[{\rm mJy}]\,(1-\beta\mu_0),
\ee 
and we have neglected the $({d\gamma }/{dt})^{abs}$ term in Eqn (\ref{eq:dp_perp_abs}) since the $({d\gamma }/{dt})$ from acceleration and from curvature and synchrotron losses are much larger.
In Eqn (\ref{eq:D}), we take $I _0 = \Phi _0\Omega_{rad} d^2/A$, 
where $\Omega_{rad} \sim A/r^2$ is 
the solid angle of radio emission, with $A$ the cross-sectional area and $r$ the radius at absorption, $\Phi _0$ is the measured radio flux (in mJy), and $d$ is the source distance (in kpc).
If and when the particles satisfy the resonance condition, $\gamma < \gamma_R$ (where $\gamma_R$ was defined in Eqn (\ref{eq:gamma_R})), as they advance along their trajectories, the resonant absorption term Eqn (\ref{eq:dp_perp_abs}) will turn on.

In the regime where $\theta_0 - \psi \ll \theta_0$, we have assumed that the pitch angle maintains a constant value of $\psi = \theta_0/2$ and that the mean of the particle total momentum is
\be
\bar p = {\Gamma({3\over 3-\nu})\over \Gamma({2\over 3-\nu})}
\left[{(3-\nu)a_1\over b_1\theta_0^2}\right]^{1/(3-\nu)}
\ee
(Petrova 2003), where   
\be
{\it a_1}(\eta ) = {4\pi ^2 {\it e}^2 {J'}_1^2(1) I_0\over c^2} 
\left( {{\varepsilon _0 \theta_0}\over {B'}mc} \right) ^{\nu },~~~~  \eta > \eta_{_R},
\label{a1}
\ee 
$b_1 = 2 e^2 B'^2/3 \hbar^2 c$ and $J'_1$ is the derivative of the Bessel function.

The radio emission is modeled as an empirical cone and core beam centered on the magnetic pole to determine the flux distribution of radio photons $S(\theta,\varepsilon_R)$, given by Eqn (9) of HSDF08, where $\theta$ is the magnetic polar angle and $\varepsilon_R$ is the radio photon energy in units of $mc^2$.  We assume this core component of the flux is emitted at a radius of  $1.8R$ and the conal component at radius (in units of the stellar radius)
\be  \label{eq:rKG}
r_{\rm KG}  \approx 40\, \left({\dot P\over 10^{ - 15}{\rm s\,s^{-1}}}\right)^{0.07} P^{0.3} \varepsilon_{GHz}^{ - 0.26} 
\ee
(Kijak \& Gil 2003), where $\varepsilon_{GHz}$ is the radio frequency in GHz.  
In order to evaluate the radio photon intensity $\Phi_0$ and the incident absorption angle $\mu_0$ at the position of the 
particle, needed in Eqns (\ref{eq:D}) and (\ref{a1}), 
we use this radio core/cone beam model.
To evaluate $\Phi_0$, we divide the radio emission beam in ovc coordinates, defined in \S \ref{sec:mag}, 
into ``beamlets" centered on tangents to the field lines and having opening angle $\mu_{bm}$.  
One set of beamlets is modulated by the cone beam at radius $r_{KG}$ (see Eqn (\ref{eq:rKG})) and another set of  beamlets is modulated by the core beam at radius $1.8 R$.  The absorption angle $\mu_0$ between each beamlet and 
a particle with which it will interact is determined by computing the travel time of the radio photons from their emission points ($x_{bm}$,$y_{bm}$,$z_{bm}$) to the particle position ($x_p$,$y_p$,$z_p$), taking into account the rotation of the field line during the photon transit time.  A particle at a given location in the outer magnetosphere can absorb only a fraction of beamlet photons.  Contributions to the cyclotron absorption from all the separate beamlets is summed at each particle position. 

 We have used this empirical model of radio core and cone emission to model the cyclotron resonant absorption, although some of the 
pulsars we study show phase alignment of their main radio and high-energy pulses, indicating that the radio emission comes from high altitude near 
the $\gamma$-ray emission.  However, the radio cone emission altitude from the Kijak \& Gil (2003) model for the Crab at 400 MHz is about 0.27 $R_{\rm LC}$  and for B1937+21 is about 0.55 $R_{\rm LC}$, which is at fairly high altitude and would likely form a caustic emission pattern to give the phase alignment with the gamma-ray pulses.   Harding et al. (2008) modeled both a cone beam at this single altitude and a cone beam with some extension along the field lines to higher altitude for the Crab in and the resulting SR was similar for both cases.  In the case of the Crab and some MSPs like B1821+24 and B1957+20 some radio peaks are in phase with the gamma-ray peaks and some are near the phase of one of the magnetic poles (as for the Crab precursor), indicating that there is both low altitude cone or core emission and high altitude radio emission simultaneously.  The cones may also be elliptical, not circular, from studies of precessing pulsars (Weisberg \& Taylor 2002).   We believe our radio model is sufficient for the present calculations since it does place the cone beam emission at high altitude for the Crab and the MSPs, in the same region with the gamma-ray emission.

 Although the electron-positron pairs from the PC cascades are thought to produce the observed radio emission, 
we have neglected any energy loss of the pairs due to emission of the radio beam.  This is reasonable since the radio emission power is a  
small fraction of the spin-down power (about $\lesssim 10^{-6}$) compared to the fraction of spin-down power used to produce the pairs 
(about $10^{-3}$, HM11).

At each step along its trajectory, the particle radiates an instantaneous synchrotron spectrum (Tademaru 1973),
\be
\dot N_{SR} (\varepsilon) = {2^{2/3}\over \Gamma({1\over 3})}\,\alpha B' \sin\psi\, \varepsilon^{-2/3}\, 
\varepsilon_{_{SR}}^{-1/3}\exp(-\varepsilon/\varepsilon_{_{SR}}),
\label{nSR}
\ee
where $\sin\psi = p_\perp/p$, $p^2 = \gamma^2 - 1$ and $\varepsilon_{_{SR}} = (3/2)\gamma^2 \, B'\sin\psi$ 
is the synchrotron critical frequency.

\subsection{Synchrotron Self-Compton Radiation} \label{sec:SSC}

Calculation of the SSC emission, the inverse-Compton radiation of both pairs and primary particles scattering the SR from both pairs and primaries, is carried out in two stages.  Since the SR photon density is needed to compute the SSC emission, the first stage calculates the SR from both pairs and primary particles at each step along all trajectories in the open field volume according to \S \ref{sec:SR}, and also accumulates the SR emissivity at each location, ${\bf r_{em}} = (x_{em}, y_{em}, z_{em})$, in Cartesian coordinates, and photon emission direction, $\boldsymbol{\eta_{em}} = (\eta_{em, x}, \eta_{em, y}, \eta_{em,z})$ in the IOF.  At each position along a particle trajectory, depending on whether the particle is a primary electron ($p$) or a pair ($+$), the SR emissivity is incremented by
\be
\Delta \epsilon_{SR} (\varepsilon, {\bf r_{em}}, \boldsymbol{ \eta_{em}}) = {\dot N_{SR}(\varepsilon)\,\Delta \ell\,\,\dot n_{p,+}
\over c \,\Delta x_{em} \Delta y_{em} \Delta z_{em}} ,
\ee
where $\Delta \ell$ is the spatial step along the particle trajectory in the IOF, $\dot n_p$ is the flux of primary electrons, given by Eqn (\ref{eq:np}), $\dot n_+$ is the flux of pairs from Eqn (\ref{eq:n+}) summed over pair energies, and $\Delta x_{em}$, $\Delta y_{em}$, $\Delta z_{em}$ are the spatial divisions in the ${\bf r_{em}}$ array.
At the end of this first stage, the SR emission and the total 
SR emissivity, $\epsilon_{SR}^N (\varepsilon, {\bf r_{em}}, \boldsymbol{\eta_{em}})$, throughout the open magnetosphere of one hemisphere (northern) has been computed.  We need the SR emissivity from both hemispheres, since particles along trajectories in one hemisphere can scatter SR photons produced by particles on trajectories in the other hemisphere, the emissivity from the southern hemisphere is computed from the one in the northern hemisphere using reflection symmetry:
\be
\epsilon_{SR}^S (\varepsilon, {\bf r_{em}}, \boldsymbol{\eta_{em}}) = \epsilon_{SR}^N (\varepsilon, -{\bf r_{em}}, -\boldsymbol{\eta_{em}}).
\ee
The total SR emissivity is then, $\epsilon_{SR} (\varepsilon, {\bf r_{em}}, \boldsymbol{\eta_{em}}) = \epsilon_{SR}^N (\varepsilon, {\bf r_{em}}, \boldsymbol{\eta_{em}}) + \epsilon_{SR}^S (\varepsilon, {\bf r_{em}}, \boldsymbol{\eta_{em}})$.

In the second stage of the simulation, the particles follow their trajectories for a second time, radiating SSC using the SR emissivity computed in the first stage.  The scattered photon distribution from a single particle is,
\be \label{eq:Nssc}
{d N_{SSC}(\varepsilon_s)\over d\varepsilon_sdt} = \int d\Omega \,\int d\epsilon\, n_{ph} (\varepsilon, \Omega)\,(1 - \beta\mu)\,{d\sigma(\varepsilon, \Omega)\over d\varepsilon d\Omega}
\ee
where $n_{ph}(\varepsilon, \Omega)$ is the SR photon density, $\sigma(\varepsilon, \Omega)$ is the scattering cross section, and $\varepsilon$ and $\varepsilon_s$ are the incident and scattered photon energies.  At each step along the particle trajectory, the SR photon density in all directions is computed at the current position, ${\bf r_{IOF}}$, of the particle,
\be
n_{ph}(\varepsilon,{\bf r_{IOF}}, \boldsymbol{\eta_{em}}) = {1\over c}\,\int_{x_{min}}^{x_{max}}\,dx_{em}\int_{y_{min}}^{y_{max}}\,dy_{em}\int_{z_{min}}^{z_{max}}\,dz_{em}\,{\epsilon_{SR} (\varepsilon, {\bf r_{em}}, \boldsymbol{\eta_{em}})\over r_s^2}
\ee
where 
\be
r_s^2 = (x_{IOF} - x_{em})^2 + (y_{IOF} - y_{em})^2 + (z_{IOF} - z_{em})^2.
\ee

In principle, the Klein-Nishina (KN) scattering cross section should be used in Eqn (\ref{eq:Nssc}).  In practice, a full integration over all angles using the full KN cross section at each particle position is not computationally feasible.  We have therefore adopted an approximate approach, using the formula of Jones (1968) for the KN photon production rate for a single ultra-relativistic particle with Lorentz factor $\gamma$ scattering an isotropic distribution, and modulated by the anisotropic photon density and the relative velocity factor:
\be \label{eq:KN}
{d N_{SSC}(\epsilon_s)\over d\varepsilon_sdt} \simeq {1\over4\pi}\int d\Omega \,\int d\epsilon\, n_{ph} (\varepsilon, {\bf r_{IOF}},\boldsymbol{\eta_{em}})\,(1 - \beta\mu)\,{dn_{KN}(\varepsilon, \epsilon_s)\over d\varepsilon_s d\varepsilon}
\ee
where

\be
{dn_{KN}(\varepsilon, \epsilon_s)\over d\varepsilon_s d\varepsilon} = {2\pi r_0^2 c\over \gamma^2\varepsilon}\,[2q\ln{q} + (1+2q)(1-q) + {1\over 2} {(\Gamma q)^2\over {1+\Gamma q}}\,(1-q)],
\ee

\be
\Gamma = 4\varepsilon\gamma, \,\, q = {E_1\over \Gamma(1-E_1)}, \,\, E_1 = {\varepsilon_s\over \gamma},
\ee
and $r_0$ is the classical electron radius, $\mu = \cos\theta$ is the incident photon angle and $\beta = (1 - 1/\gamma^2)^{1/2}$.
In the limit $\Gamma << 1$ and $E_1 << 1$, Eqn (\ref{eq:KN}) reduces to the Thompson limit result (Blumenthal \& Gould 1970).  

To compute the SSC spectral contribution from a particle at position ${\bf r_{\rm IOF}}$ with velocity ${\bf v}$ in the IOF, the integration in Eqn 
(\ref{eq:KN}) is performed by looping over the whole range of incident photon energy $\varepsilon$ and directions $\mu$ and $\phi$, relative to the particle direction.   We assume that the scattered photon direction, $\mu_s$ and $\phi_s$,  is the same as the particle direction, $\boldsymbol{\eta_{e}}$, so 
\be
\mu_s = \eta_{e,z}, \,\,\,\tan\phi_s = \left({\eta_{e,y}\over \eta_ {e,x}}\right)
\ee
To evaluate the photon density for each incident photon direction we need to know the incident photon direction, $\mu_i$ and $\phi_i$, in the IOF,
\be
\mu_i = \mu_s\mu + \sin\theta_s \sin\theta\cos\phi
\ee
and 
\be
\phi_i = \phi_s + \Delta\phi_i
\ee
with
\be
\cos\Delta\phi_i = {\cos\theta - \cos\theta_s\cos\theta_i\over \sin\theta_s\sin\theta_i}.
\ee
The quantities $\mu_i$ and $\phi_i$ define the direction $\boldsymbol{\eta_{i}}$ in the IOF that is then used to find the SR photon density in that direction.

\subsection{Particle Dynamics}

The equations of motion for the Lorentz factor, $\gamma$, and perpendicular momentum, $p_{\perp}$ (in units of $mc$) of a particle as it moves along a field line can be written (Harding, Usov \& Muslimov 2005),f
\be \label{eq:dgamma}
{d\gamma\over dt}={eE_\parallel\over mc}-{2e^4\over 3m^3c^5}
B^2\,p_\perp^2 - {{2e^2\gamma ^4}\over {3\rho _c^2}}
+ \left({d \gamma\over dt}\right)^{abs} - \left({d \gamma\over dt}\right)^{SSC} \,
\ee

\be  \label{eq:dp_perp}
{d p_\perp\over dt}=
-{3\over 2}{c\over r}{p_\perp}
-{2e^4\over 3m^3c^5}B^2\,{p_\perp^3\over \gamma} + \left({d p_\perp(\gamma)\over dt}\right)^{abs}.
\ee
In equation (\ref{eq:dgamma}), the different terms of the right hand side are acceleration, synchrotron losses,
curvature radiation losses, cyclotron/synchrotron absorption and inverse Compton losses.  
In equation (\ref{eq:dp_perp}),  the various terms of the right hand side are adiabatic momentum change along the field line, synchrotron losses and 
cyclotron/synchrotron resonant absorption.   The SSC losses are negligible for $p_\perp$.

\begin{figure}
\hskip -0.8 cm
\includegraphics[width=100mm]{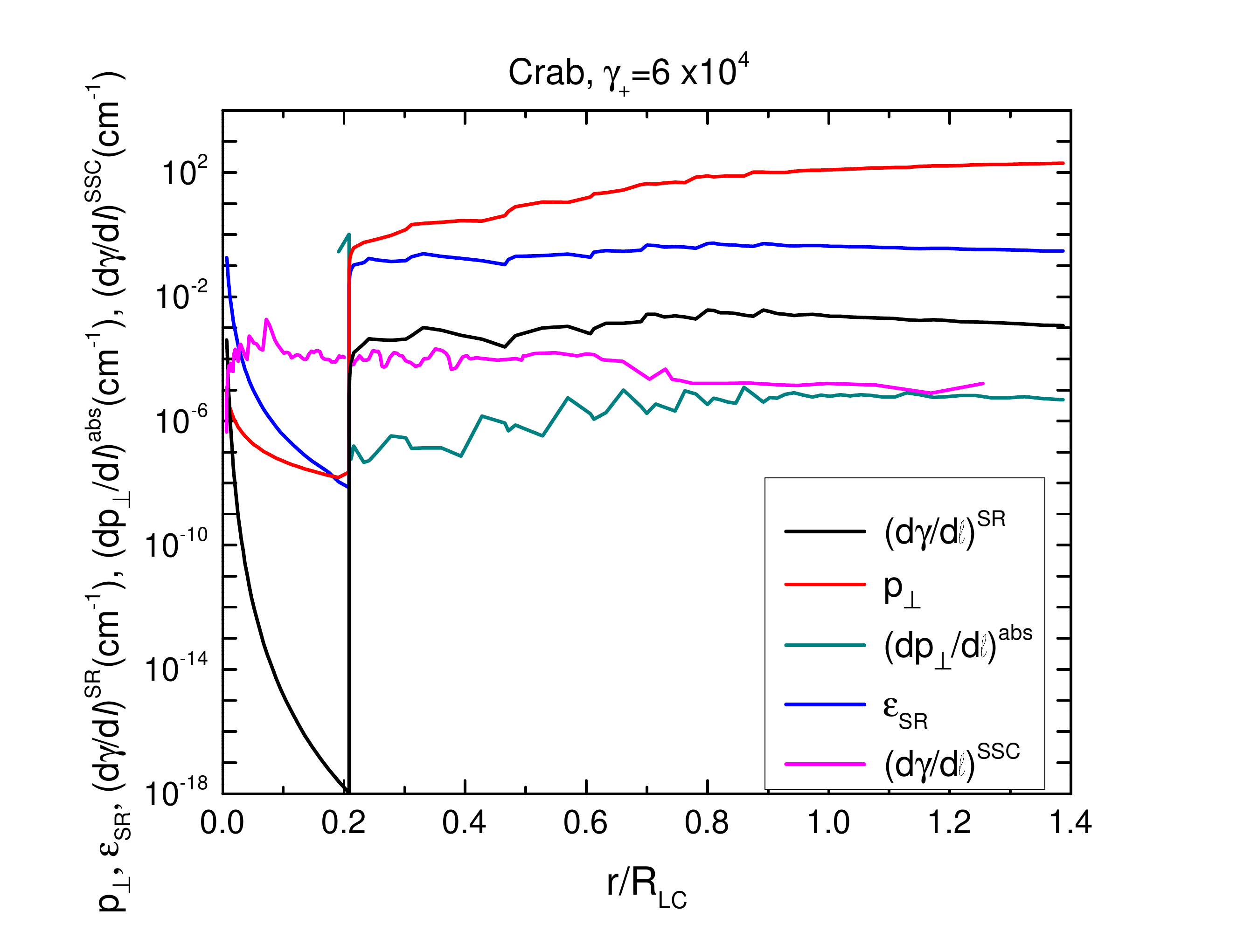}
\hskip -1.2 cm
\includegraphics[width=100mm]{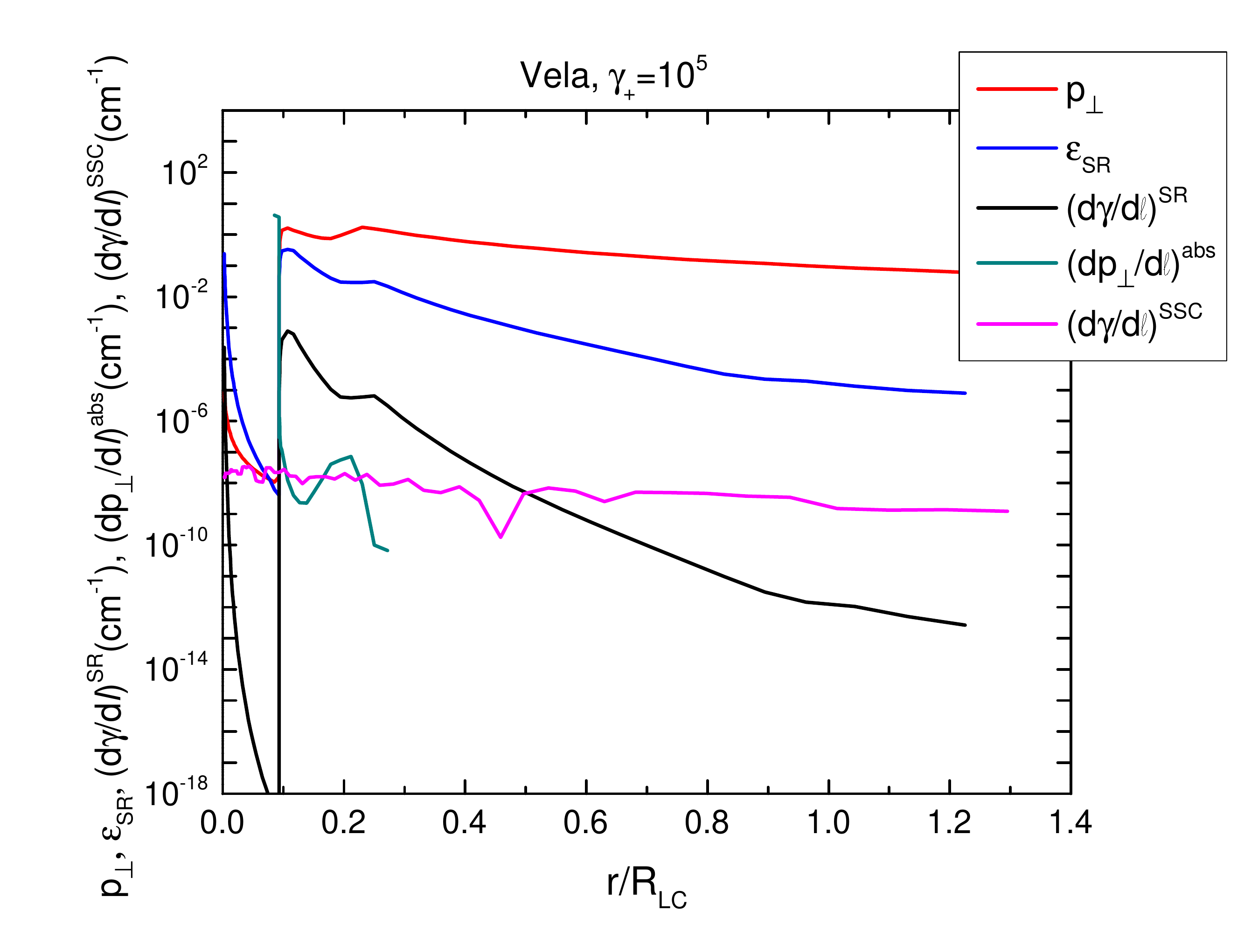}
\includegraphics[width=100mm]{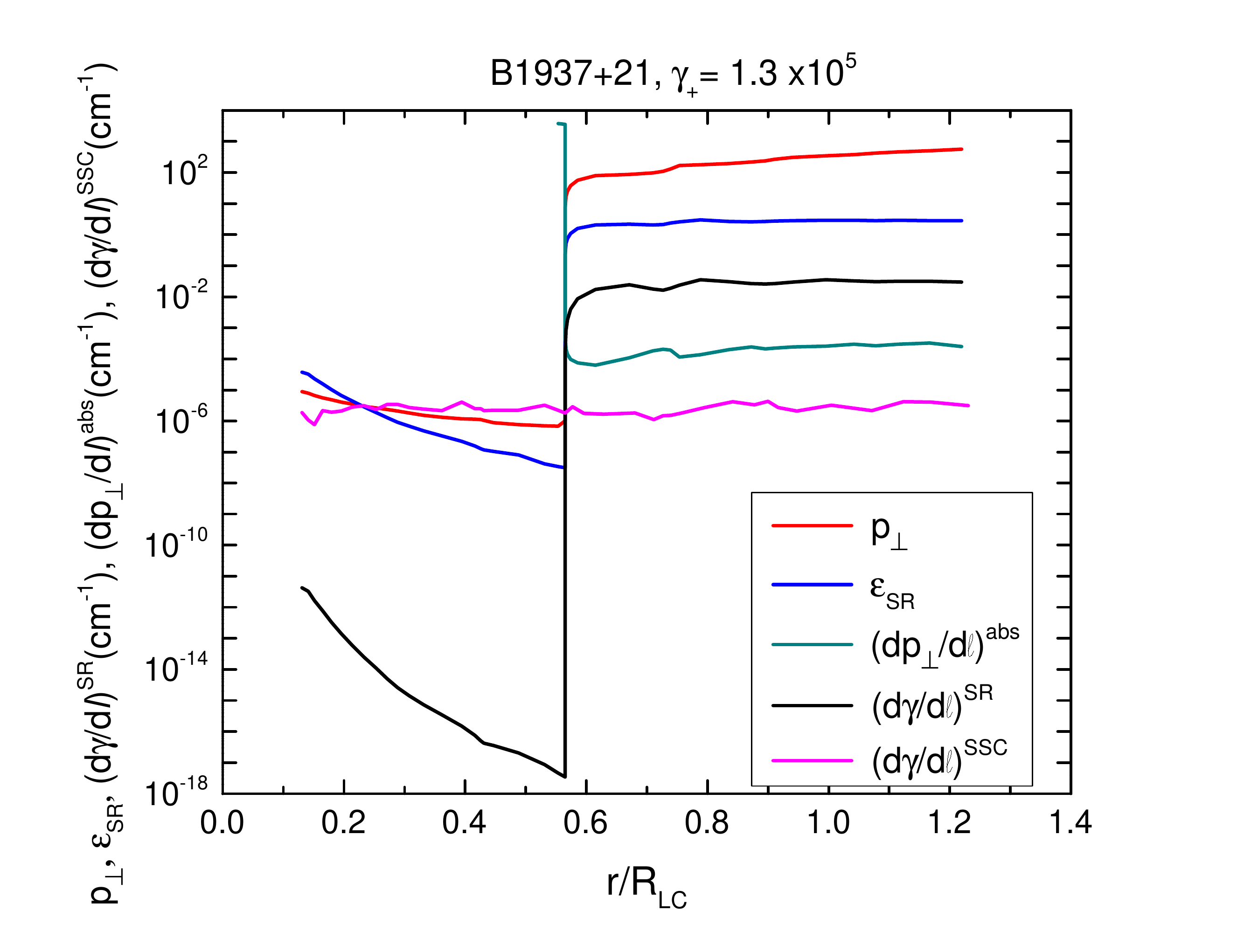}
\caption{Evolution of the dynamics and radiation of pairs at initial energy $\gamma_+$ as a function of radius (in units of light cylinder radius) 
along their trajectory, for the Crab, Vela and B1937+21.  Quantities plotted are perpendicular momentum, $p_{\perp}$ (in units of $mc$), synchrotron loss rate, $(d\gamma/d\ell)^{\rm SR}$ ($\rm cm^{-1}$), SSC loss rate, $(d\gamma/d\ell)^{\rm SSC}$ ($\rm cm^{-1}$), cyclotron absorption rate, $(dp_{\perp}/d\ell)^{\rm abs}$ ($\rm cm^{-1}$), and critical synchrotron energy $\varepsilon_{_{\rm SR}}$ (in units of $mc^2$).}    %fig. 2
 \label{fig:pair}
 \end{figure}

In the first stage, integrating the particles trajectories to compute the SR and the SR emissivity, only SR losses are included since the SSC losses are computed in the second stage.  However, we find that the SSC losses are much smaller than the SR losses in all cases except for the very high altitude part of the pair trajectories for the Vela pulsar (see Figure \ref{fig:pair} discussed below).  For this case, discussed in \S \ref{sec:results}, the SSC losses are negligible, and the SR losses do not change the particle energy significantly since the cyclotron absorption balances the SR losses.  The spatial step size for the particle trajectories is dynamically adjusted at each step to not exceed limits set by radius of curvature of the trajectory (to limit fractional change in emission direction to 10\%), acceleration rate (to limit fractional gain in primary Lorentz factor to 10\%), CR and SR loss rates (to limit fractional energy loss to 10\%), and absorption rate (to limit fractional increase in $p_\perp$ to 50\%).  

\section{Results} \label{sec:results}

The radiated spectra of primary particles and pairs have been simulated for the Crab pulsar, the Vela pulsar, and two of the most energetic  MSPs in the Northern and Southern hemispheres, PSR B1937+21 and B1821-24.  The pair spectra computed for each of the pulsars using the cascade simulation described in \S \ref{sec:pair} are used to compute the spectra of SR, CR and SSC radiation.  However, we tried several different pair multiplicities as well as the multiplicity from the original cascade simulation.  

The particle trajectories start at the neutron star surface and extend to $r_{\rm max} = 2.0 R_{\rm LC}$ in radius or $r^{\rm cyl}_{\rm max} = 1.9 R_{\rm LC}$ in cylindrical radius, whichever is reached first.  The $r^{\rm cyl}_{\rm max}$ limit is set so that the calculation stays inside the grid of the numerical field lines.  The region of footpoints of injected particles on the PCs is divided into 5 rings between $r_{\rm ovc}^{\rm min}$ and $r_{\rm ovc}^{\rm max}$, each of which is divided into 15  azimuthal divisions ($\Delta_{\rm azim} = 0.04$ divisions per degree), totaling 75 each of injected primary particles and pairs at each of 34 energies.  The code was run on 75 parallel processors on the Discover cluster at Goddard.  For the calculation of the SSC spectrum, we used 20 divisions in $\cos\theta$ and 12 divisions in $\phi$ in the integration over incident photon directions (Eqn [\ref{eq:KN}]).  The SR photon emissivity array has 50 logrithmic energy bins, from $2 \times 10^{-6} - 10^7$ MeV, 18 angular divisions in each of three independent Cartesian direction cosines, $\boldsymbol{\eta_{em}} = (\eta_{em, x}, \eta_{em, y}, \eta_{em,z})$ from $-1$ to $1$, and 9 divisions in each of three spatial coordinates ${\bf r_{em}} = (x_{em}, y_{em}, z_{em})$ from $-2.0R_{\rm LC}$ to $2.0R_{\rm LC}$.   To assure that there was enough angular accuracy in the SR photon emissivity array of the SSC computation, we tested the computed SSC flux level with increasing number of angular divisions.  It was found that the SSC flux decreases with increasing number of divisions but stabilizes at 18 divisions in each component of $\boldsymbol{\eta_{em}}$, for a combination of $18^3 = 5832$ elements over $4\pi$ steradians, so this number was adopted.

Figure \ref{fig:pair} shows examples of the evolution of the pair perpendicular momentum, $p_\perp$, cyclotron absorption rate, $(d p_\perp /d\ell)^{abs}$, synchrotron loss rate, $(d\gamma/d\ell)^{SR}$, SSC loss rate, $(d\gamma/d\ell)^{SSC}$, and synchrotron critical energy, $\varepsilon_{SR}$, as a function of radial distance along a pair trajectory for the Crab, Vela and B1937+21, from integration of Eqns  (\ref{eq:dgamma}) and (\ref{eq:dp_perp}), all for pair energies around $\gamma_+ \sim 10^5$.  The cyclotron absorption starts at the radius of the radio cone emission, which is about $0.2 R_{\rm LC}$ for the Crab, about $0.1 R_{\rm LC}$ for Vela, and $0.56 R_{\rm LC}$ for B1937+21, according to Eqn (\ref{eq:rKG}).   Before the radio emission altitude is reached, there is some transient SR from having to set a small non-zero initial $p_\perp = 10^{-5}$, to avoid overflow in evaluation of $(d p_\perp /d\ell)^{abs}$ in Eqn (\ref{eq:dp_perp_abs}).  The fluctuations in the absorption rate as a function of radius are due to emission from different beamlets the electron encounters (see \S \ref{sec:SR}), reflecting the numerical resolution of these sub-elements of the radio cone beam.  For this pair energy, the $p_\perp$ becomes relativistic.
In the case of the Crab and B1937+21, the radio photons are emitted at high enough altitude and the magnetic field strength near the light cylinder, $B_{\rm LC}$, is high enough that the radio photons stay in resonance in the particle rest frame through the entire particle trajectory.  Their Lorentz factors (not plotted) are nearly constant since the cyclotron absorption rate comes into equilibrium with the SR loss rate.  In the case of Vela, the absorption occurs over only a small part of the trajectory above the radio emission altitude, above which the particle is no longer ``in sight" of the relatively more narrow radio beam.  The $p_\perp$, $(d\gamma/d\ell)^{SR}$  and $\varepsilon_{SR}$ thus drop with increasing radius.  The SSC loss rate is always less than the SR loss rate, except in the case of Vela where the SR loss rate drops below the SSC loss rate at higher altitude.  
 The SSC loss rate for Vela does not decrease much at high altitude.  Because the $B_{\rm LC}$ for Vela is much lower compared to the other pulsars, the SR critical frequency is about four order of magnitude lower, so that the radiated photon number density increases with altitude.
Although the SSC loss rates are not included in the trajectory integration, they are never large enough to change the particle energy significantly.  
 
\begin{figure}
\hskip -0.8 cm
\includegraphics[width=100mm]{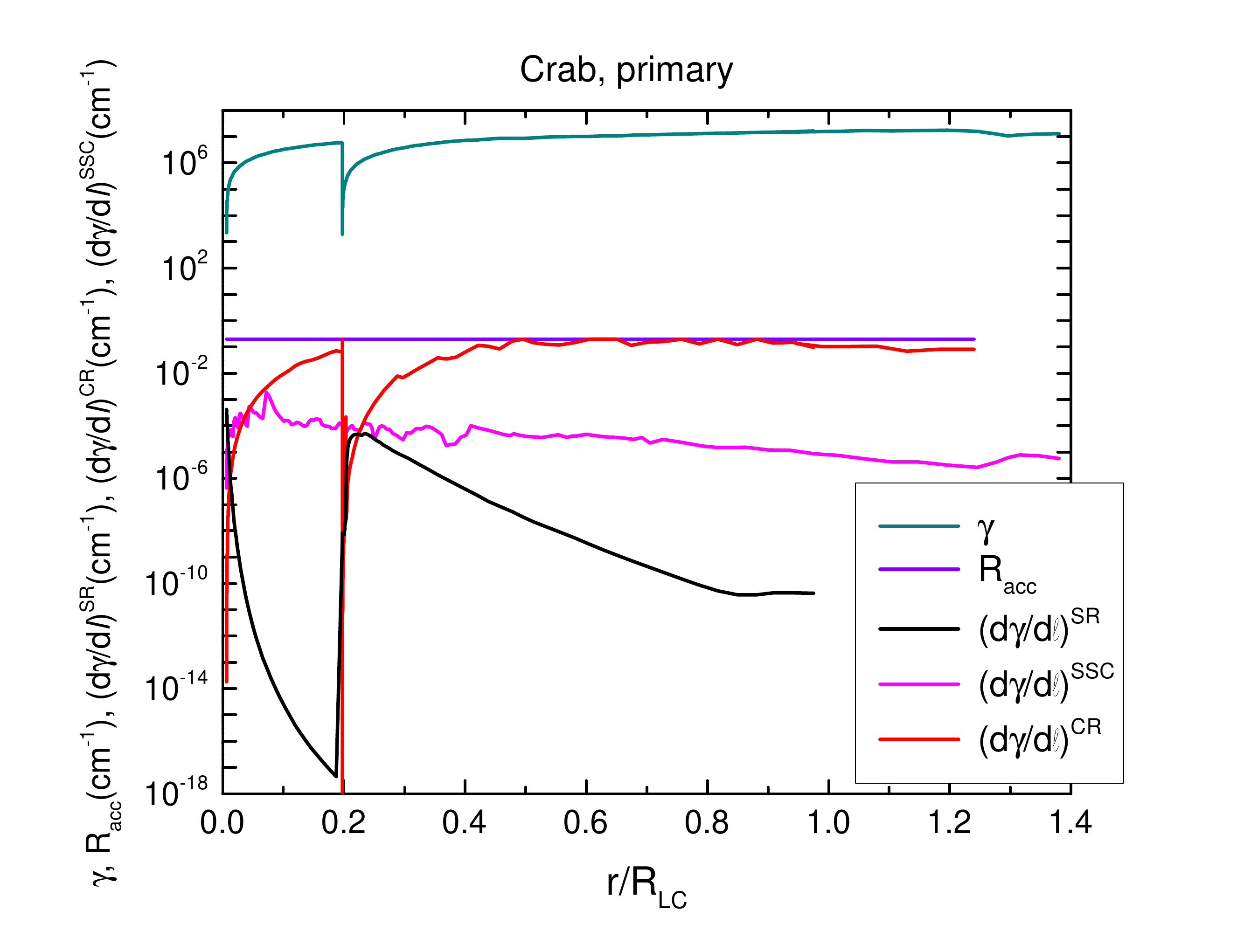}
\hskip -0.5 cm
\includegraphics[width=100mm]{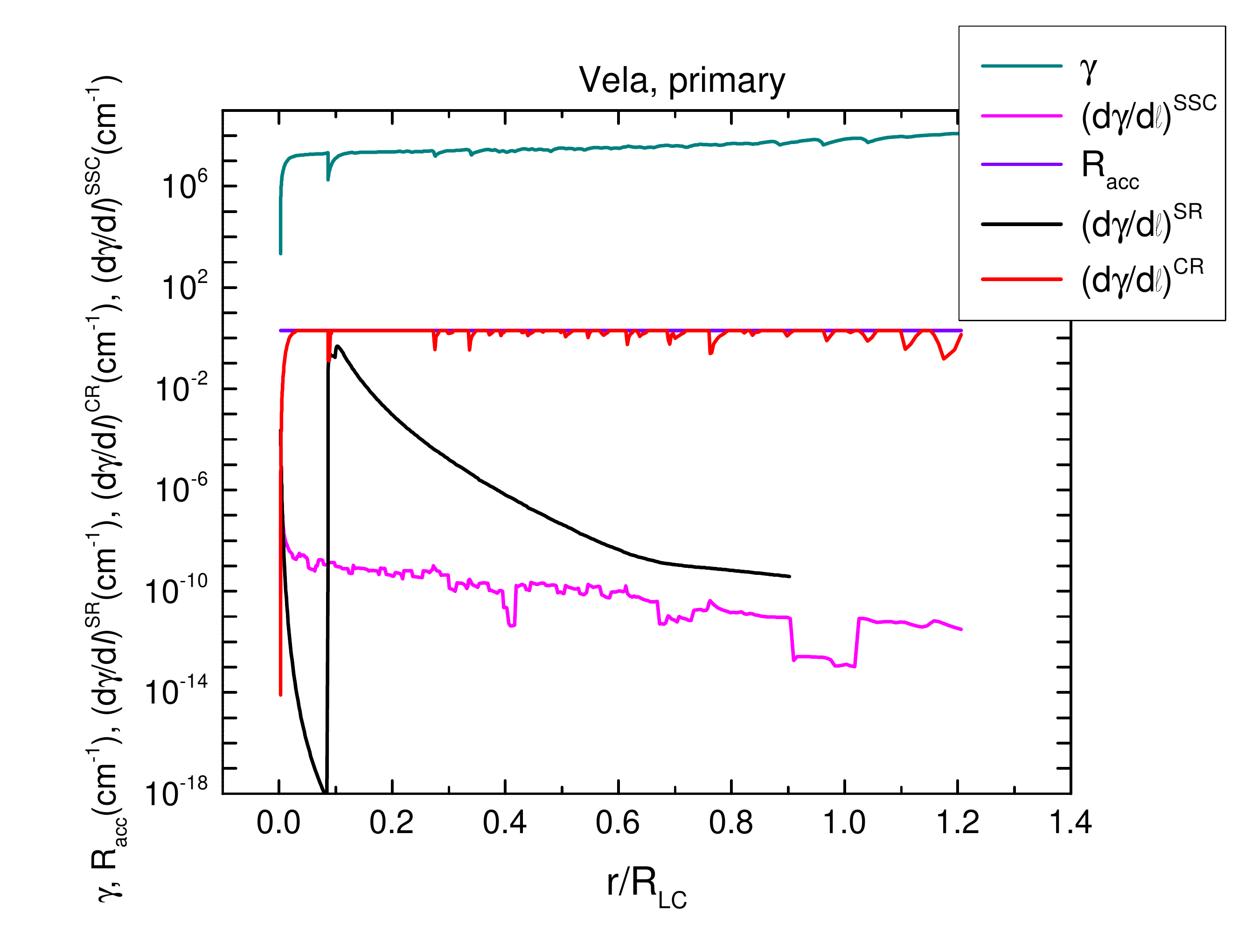}
\includegraphics[width=100mm]{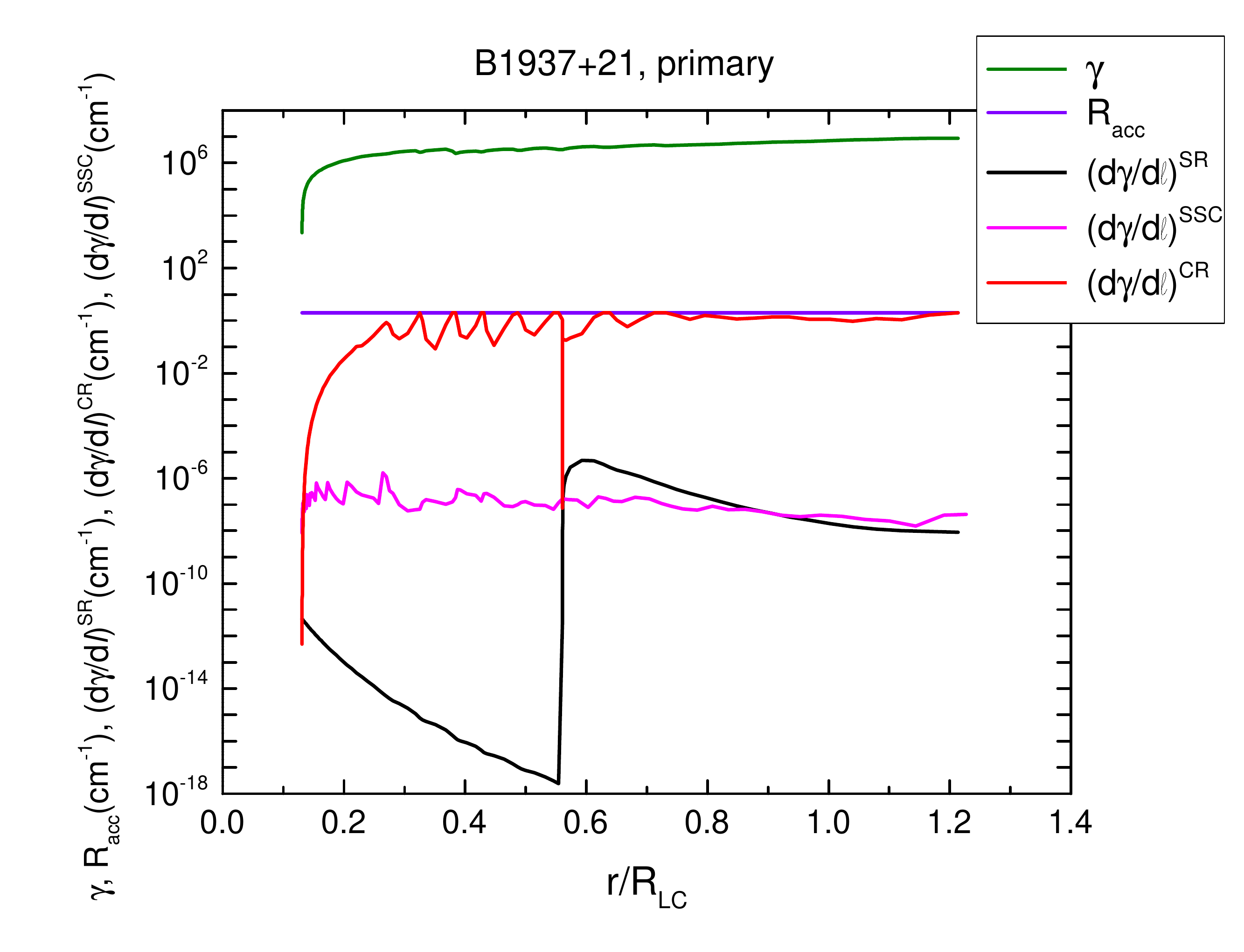}
\caption{Evolution of the dynamics and radiation of primary electrons as a function of radius (in units of light cylinder radius) 
along their trajectory, for the Crab, Vela and B1937+21.  Quantities plotted are Lorentz factor, $\gamma$, synchrotron loss rate, $(d\gamma/d\ell)^{\rm SR}$ ($\rm cm^{-1}$), CR loss rate, $(d\gamma/d\ell)^{\rm CR}$ ($\rm cm^{-1}$), SSC loss rate, $(d\gamma/d\ell)^{\rm SSC}$ ($\rm cm^{-1}$), and acceleration gain rate, $R_{\rm acc}$ ($\rm cm^{-1}$).}    %fig. 3
 \label{fig:primel}
 \end{figure}

Figure \ref{fig:primel} shows examples of the evolution of the primary electron Lorentz factor $\gamma$, curvature radiation loss rate, $(d\gamma/d\ell)^{CR}$, synchrotron loss rate, $(d\gamma/d\ell)^{SR}$ and SSC loss rate, $(d\gamma/d\ell)^{SSC}$, as a function of radial distance along a pair  trajectory for the Crab, Vela and B1937+21.  For the primaries, CR losses are always much larger than SR and SSC losses and the electrons reach CR reaction limit in all cases within a few tenths of $R_{\rm LC}$.  The maximum Lorentz factors for the chosen values of acceleration rate are slightly over $10^7$, giving critical CR energies of a few GeV.  The SR loss rate peaks soon after the start of the cyclotron resonant absorption, at which point the particle Lorentz factor drops as the SR loss and acceleration gain rates are briefly in equilibrium (Harding et al. 2005).  As the SR loss rate drops, the particle then returns to equilibrium between acceleration gain and CR loss rates.  

\subsection{Crab Pulsar}

For the Crab pulsar, we use the following measured parameters: $P = 0.033$ s, $\dot P = 4.22 \times 10^{-13}\,\rm s\,s^{-1}$, $d = 2$ kpc and $S400 = 700$ mJy, where $d$ is the distance and $S400$ is the radio flux at 400 MHz.  In addition, we assumed a magnetic inclination angle of $\alpha = 45^\circ$, a constant acceleration rate of $R_{\rm acc} = eE_\parallel / mc^2 = 0.2\,\rm cm^{-1} = 6 \times 10^{-4}\,B_{\rm LC}$, where $B_{\rm LC} = 6 \times 10^5$ G, for the primary particles, and two pair multiplicities, $M_+ = 2 \times 10^4$, from a steady-state pair cascade, and $M_+ = 3 \times 10^5$, from a time-dependent pair cascade.

\begin{figure}
\includegraphics[width=180mm]{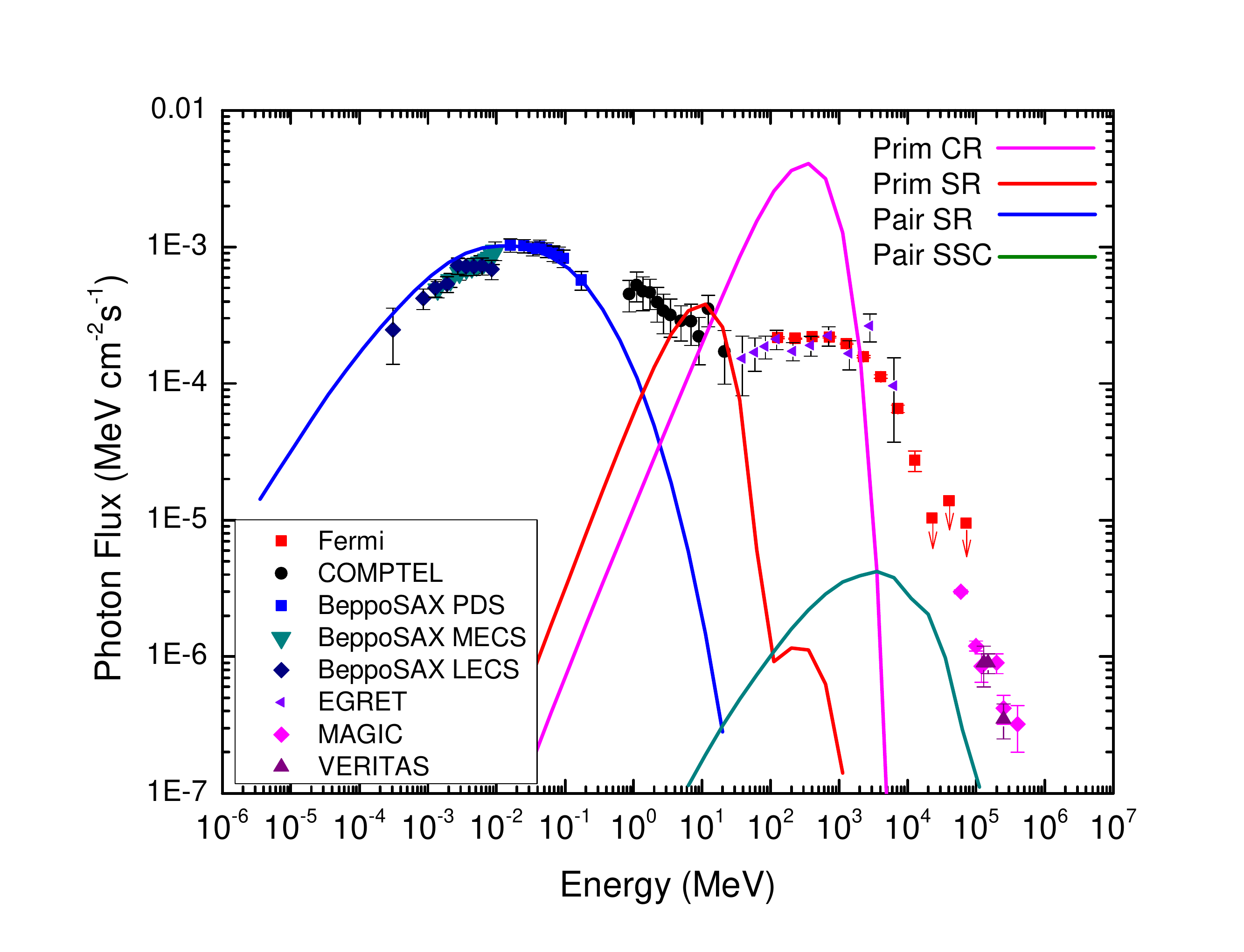}
\caption{Model spectra of phase-averaged pulsed emission components from primary electrons and pairs (as labeled) from the Crab pulsar, for magnetic inclination angle $\alpha = 45^\circ$ and observer angle $\zeta = 60^\circ$ and pair multiplicity $M_+ = 2 \times 10^4$.  Data points are from Kuiper et al. (2001) [http://www.sron.nl/divisions/hea/kuiper/data.html], Abdo et al. (2013) [http://fermi.gsfc.nasa.gov/ssc/data/access/lat/2nd\_PSR\_catalog/], Aleksic et al. (2012) and Aliu et al. (2011).}    %fig. 4
 \label{fig:Crab1}
 \end{figure}

Figure \ref{fig:Crab1} shows the observed and modeled spectral energy distribution (SED) of the phase-averaged flux of the Crab pulsar from optical to VHE $\gamma$-ray energies at viewing angle $\zeta = 60^\circ$ and for $M_+ = 2 \times 10^4$.  The different model radiation components, primary CR, SR, SSC and pair SR and SSC, are plotted separately.  We multiplied the phase-averaged pair SR flux by a factor of 11 to match the observed spectral points and multiplied all the other components by the same factor.  The shape of the pair SR component, dependent on the shape of the pair spectrum, and its peak energy, dependent on the magnetic field strength near the light cylinder, nicely matches the observed spectrum from optical through the BeppoSax hard X-ray measurements.  However, it falls short of the COMPTEL low-energy $\gamma$-ray points.  The pair SSC component peaks at several GeV but falls well below the observed {\it Fermi} and VHE spectral points.  The primary CR component peaks near a GeV but way above the observed spectrum, while the lower primary SR component peaks around 1 MeV at the level of the observed spectrum.  The primary SSC component, peaking sharply at a few times $10^6$ MeV, has a flux below the scale of the plot.

\begin{figure} 
\includegraphics[width=180mm]{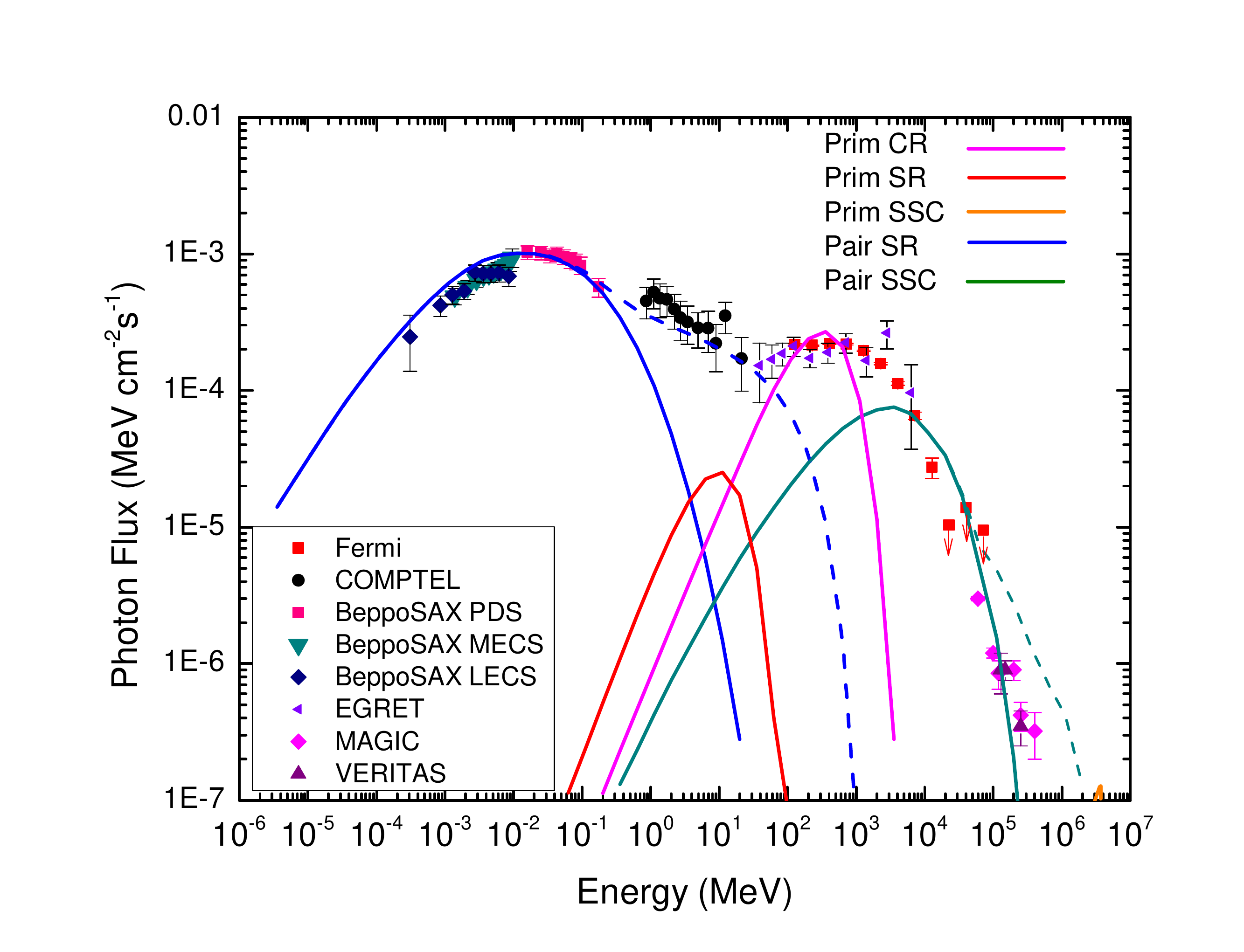}
\caption{Model spectra of phase-averaged pulsed emission components from primary electrons and pairs (as labeled) from the Crab pulsar, for magnetic inclination angle $\alpha = 45^\circ$ and observer angle $\zeta = 60^\circ$ and pair multiplicity $M_+ = 3 \times 10^5$.   The dashed lines are the SR and SSC spectra resulting from a power law extension to the cascade pair spectrum.}    %fig. 5
\label{fig:Crab2}
\end{figure}

Figure \ref{fig:Crab2} shows the observed and modeled SED for the same parameters as the model in Figure \ref{fig:Crab1} but for a higher pair multiplicity of $M_+ = 3 \times 10^5$.  For this case, the pair SR spectrum matches the observed spectrum without any renormalization factor.  The associated pair SSC component now roughly matches the level of the higher-energy {\it Fermi} points and the MAGIC and VERITAS measurements.  For the adopted value of the acceleration rate $R_{\rm acc} $, the primary CR component fills in at a few hundred MeV.  However, the primary SR component is too low to account for the observed soft $\gamma$-ray emission.  The tip of the steeply rising, nearly mono-energetic primary SSC spectrum is just visible at a few TeV at the bottom right edge of the plot, and the pair CR lies  several orders of magnitude below the plot lower boundary.  

We tested the addition of a high-energy power law extension to the pair spectrum (shown in Figure \ref{fig:Crab2}) whose SR spectrum would account for the observed emission in the 1 - 100 MeV range.  The resulting pair SR and SSC opponents are shown in Figure \ref{fig:Crab2} as dashed lines.  The SSC spectrum now extends to higher energy with a harder spectrum that exceeds the observed MAGIC and VERITAS points.  This extension of the pair spectrum, which has no physical basis, would thus produce an SSC spectrum that seems to be ruled out by the data.  This implies that the observed 1-10 MeV emission is not produced by the same particles that produce the SSC emission.

\subsection{Vela Pulsar} \label{sec:Vela}

For the Vela pulsar, we use the following measured parameters: $P = 0.089$ s, $\dot P = 1.25 \times 10^{-13}\,\rm s\,s^{-1}$, $d = 0.29$ kpc and $S400 = 5000$ mJy.  We assumed a magnetic inclination angle of $\alpha = 75^\circ$, a constant acceleration rate of $R_{\rm acc} = eE_\parallel / mc^2 = 2.0 = 0.08\,B_{\rm LC}$ for the primary particles, and two pair multiplicities, $M_+ = 6 \times 10^3$, from the steady-state pair cascade, and $M_+ = 1 \times 10^5$, for a time-dependent pair cascade.  

\begin{figure} 
\includegraphics[width=180mm]{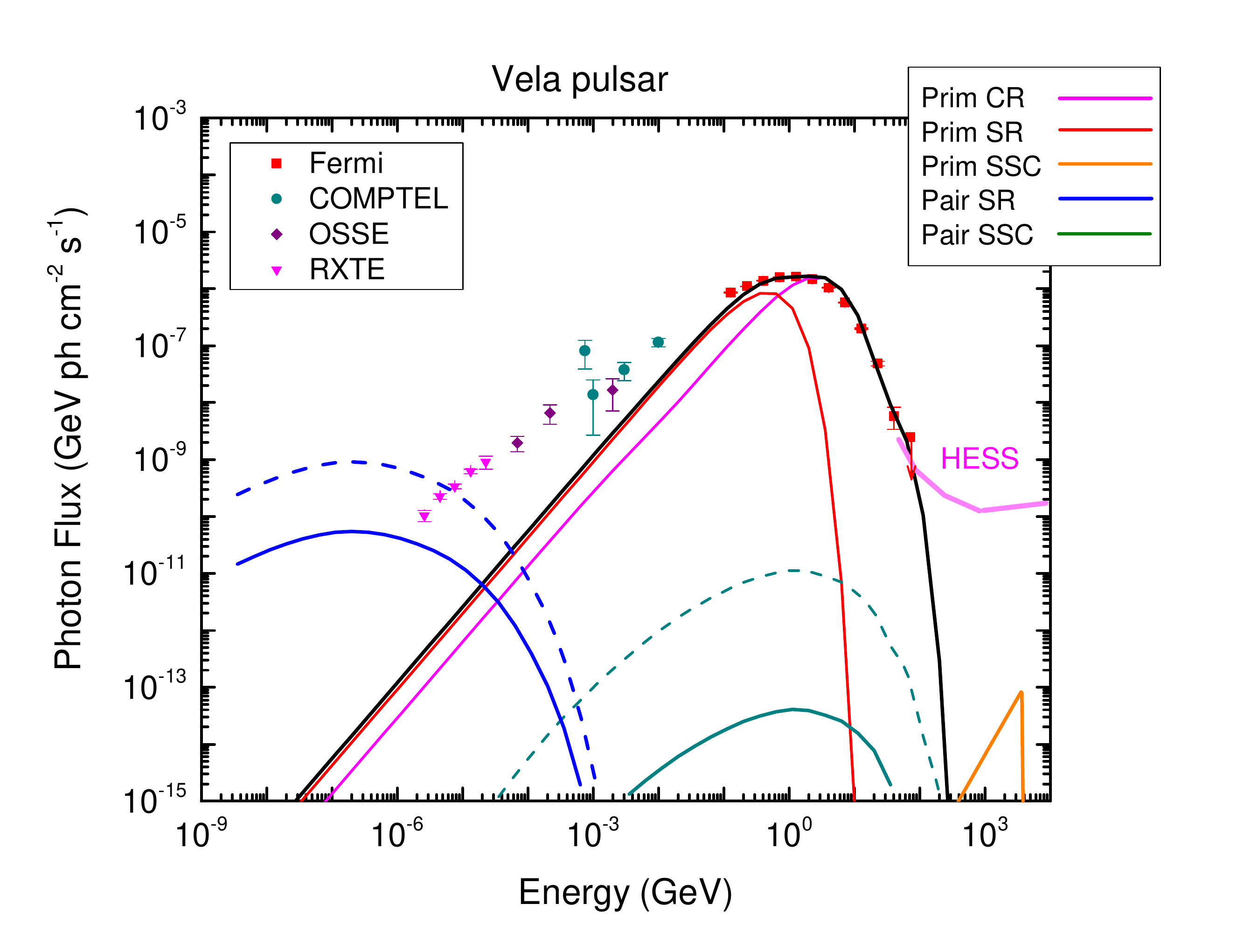}
\caption{Model spectra of phase-averaged pulsed emission components from primary electrons and pairs (as labeled) from the Vela pulsar, for magnetic inclination angle $\alpha = 75^\circ$ and observer angle $\zeta = 60^\circ$.    Solid lines are for pair multiplicity $M_+ = 6 \times 10^3$ and dashed lines are for $M_+ = 10^5$.  Data points are from Abdo et al. (2013) [http://fermi.gsfc.nasa.gov/ssc/data/access/lat/2nd\_PSR\_catalog/] and Harding et al. (2002).}    %fig. 6
\label{fig:Vela}
\end{figure}

Figure \ref{fig:Vela} shows the observed and models SED of the phase-averaged flux of Vela at viewing angle $\zeta = 60^\circ$ for the two different values of $M_+$.  We multiplied the phase-averaged primary CR flux by a factor of 0.14 to match the level of the observed {\it Fermi} spectral points and multiplied all the other components by the same factor.  The value of $R_{\rm acc}$ had been chosen so that the SED peak of the primary CR spectrum matches the peak of the {\it Fermi} spectrum, but the flux level being too high implies that either the flux of primary particles or the distance over which they are radiating is smaller than what we assume.  The pair SR and SSC components peak at energies of around 1 keV and 1 GeV respectively, somewhat lower than the peaks of those components for the Crab.  For both values of $M_+$, the SSC flux is orders of magnitude lower than the primary CR flux.  The primary SSC component is well below observable levels.  The striking differences between the Vela and Crab SEDs come from a number of differences in their intrinsic properties.  First, although they have similar surface magnetic field strengths, the larger magnetosphere of Vela gives it a much lower $B_{\rm LC} = 4 \times 10^4$ G compared to that of the Crab.  This produces the lower peak energies of the SR and SSC components and also the lower flux levels because the magnetic field drops below the value where the radio frequencies are in the cyclotron resonance in the pairs' rest frames well before the light cylinder, whereas for the Crab, the radio photons are in resonance out to and beyond the LC.  Second, the radio luminosity of Vela is lower by about a factor of 5 compared to the Crab, and is produced at lower altitude relative to the LC.  The particles therefore see a lower density of radio photons over their trajectories and undergo less absorption. Third, the pair spectrum of Vela has a high-energy turnover at a lower energy, producing the lower peak energies of SR and SSC spectra.  

\subsection{Millisecond Pulsars}

MSPs are promising sources of high non-thermal SR and SSC emission, since they have small magnetospheres and consequently a number have large $B_{\rm LC}$.  In fact, a larger fraction of MSPs have observed hard non-thermal X-ray spectra that do non-recycled pulsars.  We model two of the most energetic MSPs, B1821-24 and B1937+21, that have the highest $B_{\rm LC}$ and high levels of non-thermal X-ray emission.   B1821-24 (J1824-2452), with $P = 3.05$ ms, $\dot P = 1.6 \times 10^{-18}\,\rm  s\, s^{-1}$ and $S400 = 40$ mJy, lies in the globular cluster M28, at a distance of $d = 5.5$ kpc.  B1937+21 (J1939+2134), with $P = 1.6$ ms, $\dot P = 1.0 \times 10^{-19}\,\rm s\, s^{-1}$, $S400 = 240$ mJy and $d = 3.6$ kpc, was the first discovered MSP (Backer et al. 1982).  Both are isolated MSPs, have $B_{\rm LC} = 7.3 \times 10^5$ G and $10^6$ G respectively, exhibit giant radio pulses (Bilous et al. 2015, Popov \& Stappers 2003) and have aligned high-energy and radio profile components similar to the Crab.  

\begin{figure} 
\includegraphics[width=180mm]{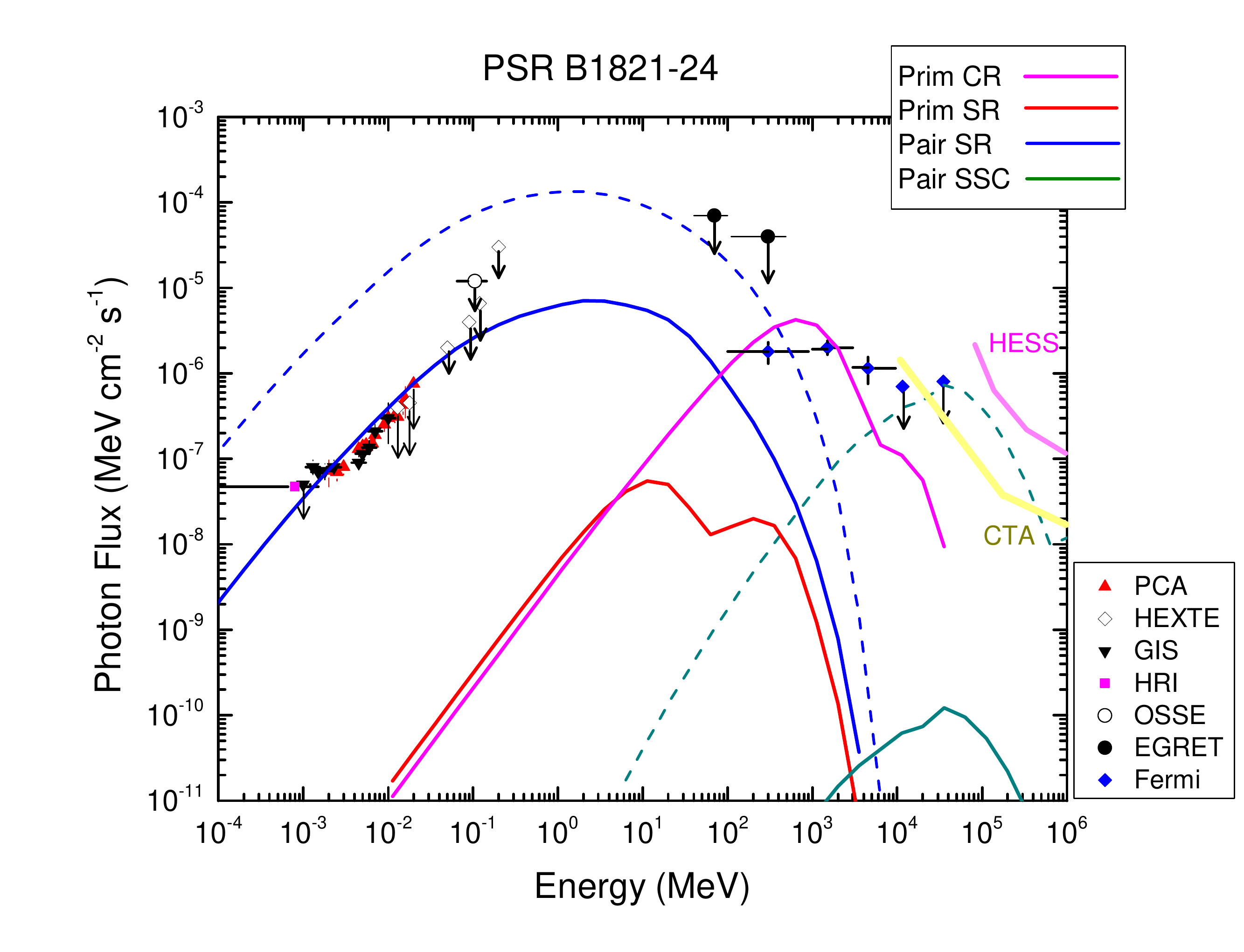}
\caption{Model spectra of phase-averaged pulsed emission components from primary electrons and pairs (as labeled) from PSR B1821-24, for magnetic inclination angle $\alpha = 45^\circ$ and observer angle $\zeta = 80^\circ$.  Solid lines are for pair multiplicity $M_+ = 10^3$ and dashed lines are for $M_+ = 10^5$.  Data points are from Kuiper et al. (2005) and Johnson et al. (2014).  The thick pink and yellow lines are the HESS and CTA sensitivity limits, respectively.}    %fig. 7
\label{fig:B1821}
\end{figure}

Figure \ref{fig:B1821} shows the observed and modeled spectra of B1821-24 from soft X-rays through VHE $\gamma$-ray energies for $\alpha = 45^\circ$ and viewing angle $\zeta = 80^\circ$.  Although there are no strong constraints on either $\alpha$ or $\zeta$, modeling the $\gamma$-ray light curve only favors $\alpha \sim 40^\circ$ and $\zeta = 85^\circ$ (Johnson et al. 2013).  We compute SR and SSC spectra for $M_+ = 10^3$, a value for a pair cascade with offset PC of $\epsilon = 0.6$, and also an extreme value of $M_+ = 10^5$, both assuming $R_{\rm acc} = eE_\parallel / mc^2 = 2.0 = 5 \times 10^{-3}\,B_{\rm LC} $ for the primary particles.  The pair SR component peaks between 1 and 10 MeV and the pair SSC component peaks around 50 GeV, almost two decades higher than the SR and SSC components in the Crab SED.  This is expected, given than the pair spectrum of B1821-24 is higher in energy by a similar factor relative to the Crab pair spectrum, and the $B_{\rm LC}$ values are similar.  The pair SR spectrum for the $M_+ = 10^3$ case matches the slope and level of the X-ray data with a minor shift by a factor of 0.8, which has been applied to the other components for this model.  The primary CR component roughly coincides with the {\it Fermi} points, with the primary SR being much lower.  The pair SSC spectrum lies about four decades below the pair SR and primary CR components and is not detectable.  The pair SR flux for $M_+ = 10^5$ is almost two decades above the observed X-ray flux, although the pair SSC flux might be detectable with {\it Fermi} and even HESS.  Since time-dependent cascade simulations for MSPs are not available at present, we cannot say whether such high multiplies are in fact achievable, although a pair multiplicity much higher than $M_+ = 10^3$, which already accounts for the observed X-ray spectrum, seems to be ruled out.  Therefore, for our chosen values of inclination and viewing angle, we do not predict a detectable VHE component from B1821-24.

\begin{figure} 
\includegraphics[width=180mm]{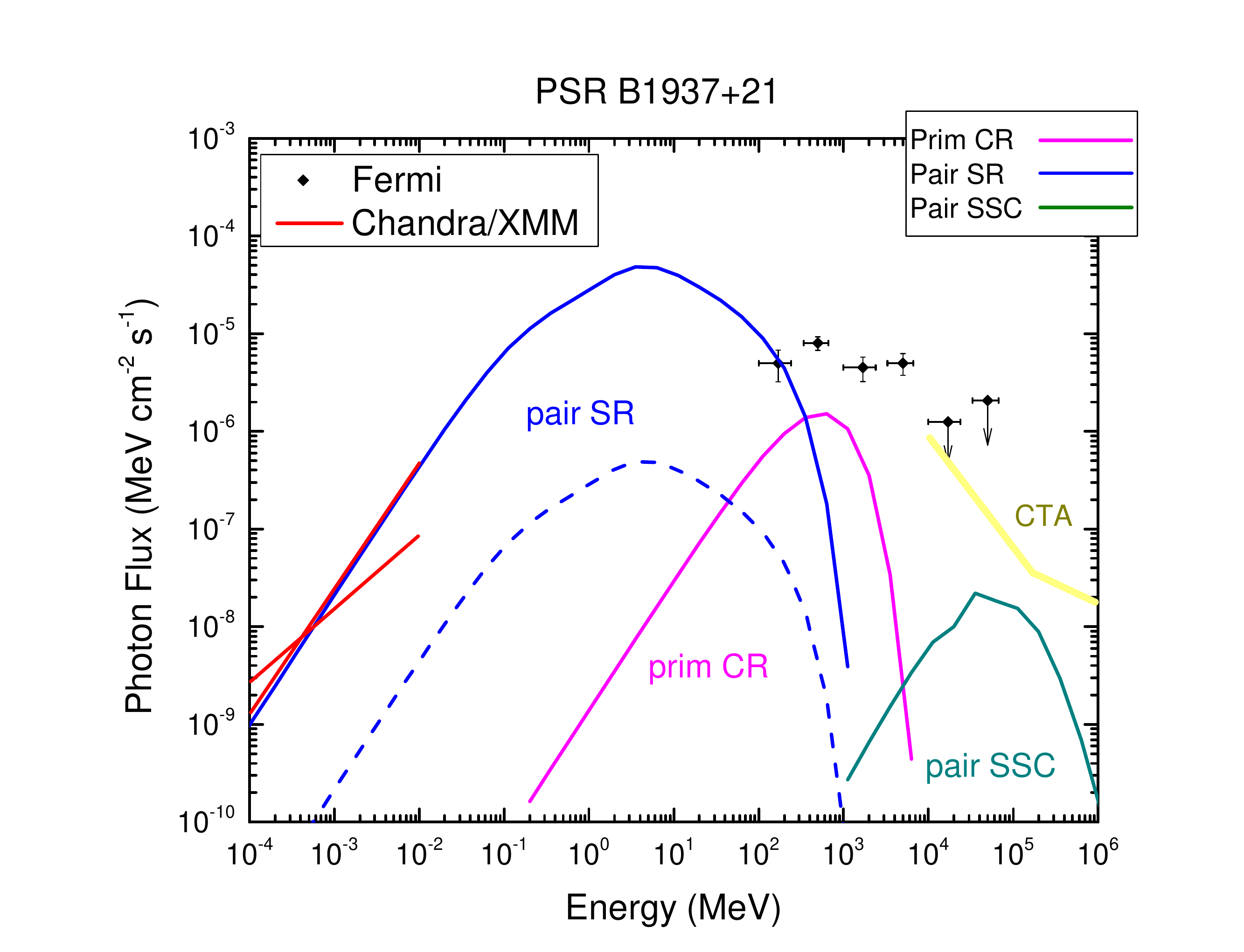}
\caption{Model spectra of phase-averaged pulsed emission components from primary electrons and pairs (as labeled) from PSR B1937+21, for magnetic inclination angle $\alpha = 75^\circ$ and observer angle $\zeta = 70^\circ$.  Solid lines are for pair multiplicity $M_+ = 10^5$ and dashed lines are for $M_+ = 10^3$.  Data points are from Ng et al. (2014) and Guillemot et al. (2012).  The thick yellow line is the CTA sensitivity limit.}    %fig. 8
\label{fig:B1937}
\end{figure}

Figure \ref{fig:B1937} shows the observed and modeled spectra of B1937+21 from soft X-rays through VHE $\gamma$-ray energies for $\alpha = 75^\circ$ and viewing angle $\zeta = 70^\circ$.   Johnson et al. (2014) fit the combined $\gamma$-ray and radio light curves of this MSP using several geometrical emission models.   For the two-pole caustic geometry, they find $\alpha = 88^\circ (+2-1)^\circ, \zeta = 88^\circ (+2-1)^\circ$, and for an OG geometry, $\alpha=72^\circ \pm 1^\circ, \zeta=85^\circ \pm 1^\circ$.  We show the spectra for pair multiplicities of $M_+ = 10^3$ of a pair cascade with $\epsilon = 0.6$, and also for $M_+ = 10^5$, both assuming $R_{\rm acc} =  2.0 = 3 \times 10^{-3}\,B_{\rm LC} $ for the primary particles.  Applying only the $1/(2\pi d^2)$ factor to obtain the phase-averaged flux from the sky map, the pair SR flux for $M_+ = 10^5$ is consistent with the Chandra X-ray flux.  The peak of the pair SR SED is around 5 MeV, very similar to the SR SED of B1821-24.  The primary CR component is about a factor of 3 below the {\it Fermi} flux points, and the pair SSC component, peaking at around 100 GeV, is about a factor of 2 below CTA sensitivity.  The pair SR component for the case of $M_+ = 10^3$ is about two decades below the observed X-ray flux and the pair SSC emission lies well below the plot boundary.  The SR fluxes of  B1821-24 and B1937+21 for the same pair multiplicity and similar are very different since the model inclination angles are quite different.  The value of $\alpha = 75^\circ$ was chosen to match the best model fits of the $\gamma$-ray light curve, but a simulation for B1937+21 with $\alpha = 45^\circ$ would  match the observed X-ray spectrum with a lower pair multiplicity.   In this case, the SSC component would be much lower.

\section{Discussion}

We have simulated the synchrotron and synchrotron-self Compton emission from a broad spectrum of electron-positron pairs in a global force-free magnetosphere.  The scattering takes place in the extreme Klein-Nishina limit and although the angular dependence of the cross section is neglected, the full 3D angular dependence of the SR photon density is treated by storing the SR emissivity over the entire magnetosphere from both rotational hemispheres.  

Our simulation of the Crab pulsar radiation reproduces both the flux level and the shape of the observed optical to hard X-ray emission assuming a pair multiplicity of $M_+ = 3 \times 10^5$.  Such a high multiplicity is about an order of magnitude larger than that produced in steady-state PC cascades, but is achievable with the more realistic time-dependent cascades (Timokhin \& Harding 2015).  The predicted SSC flux in this case roughly matches the tail of the observed {\it Fermi} spectrum as well as the MAGIC and VERITAS detected points.  A CR component from primary particles is necessary to explain the {\it Fermi} spectrum below a few GeV.  The model SSC spectrum is not a power law, but reflects the shape of the pair spectrum, which has a smooth and continuous curvature extending to $\sim 1$ TeV.  An SSC component from primary particles appears above a few TeV and may ultimately be detectable.

The Vela pulsar does not produce detectable SSC emission in our simulations, even for a pair multiplicity of $10^5$.  The predicted SSC emission has both a lower flux and a lower SED peak energy than the Crab SSC emission, because the SED peak energy of the Vela pair SR is several decades lower.  This difference is due to a combination of lower pair energy and lower $B_{\rm LC}$.  However, the {\it Fermi} spectrum can be produced by primary CR and SR.  

Simulating emission from two of the most energetic MSPs, B1821-24 and B1937+21, we find that their observed X-ray emission is well reproduced by SR from pairs.  The SEDs of MSP SR components are predicted to peak at 1 - 10 MeV and could be detected by proposed Compton telescopes that would test this model.   Since the soft X-ray emission is produced by the pairs at the low-energy turnover in the pair spectrum, well below the SED peak, the spectral index is that expected for SR from a single particle, $-2/3$ (4/3 for the SED).  This index is much higher than the soft X-ray index  of the pair SR spectrum of the Crab, where the soft X-ray emission is near the SED peak and produced by a range of pair energies.  However, the predicted SSC fluxes of the MSPs are too low to be detectable with current instruments.  

Thus from our simulations, only Crab and Crab-like pulsars, such as B1509-58 and B0540-69 in the LMC, are expected to have high enough levels of SSC emission at VHE energies to possibly be detectable by ground-based Air-Cherenkov telescopes.  They share the characteristics that are important for production of strong SSC emission: high levels of non-thermal X rays, high pair multiplicity and high magnetic fields near the light cylinder.  
Non-Crab-like, middle-aged pulsars like Vela have much lower levels of non-thermal X rays relative to their GeV emission.  So even if they are able to generate very high pair multiplicities $\sim 10^5$, their pair energies will be too low and their SSC emission will not reach detectable levels at VHE energy.  
MSPs have higher levels of SSC than middle-aged pulsars, and their SSC spectra extend to higher energies than Crab-like pulsars, since their pair spectra reach TeV energies.  However, their pair multiplicities may not be high enough to generate fluxes of SSC emission that are detectable by current telescopes.  These findings are consistent with the recent stacking analysis of McCann (2015).

Lyutikov (2013) presented a model for the Crab pulsar emission, assuming cyclotron and cyclotron-self Compton emission by pairs produced in the OG and gyrating at the cyclotron resonance with small pitch angles.  It was suggested that the  pairs acquire their pitch angles through coherent emission of radio waves, but their perpendicular momenta are assumed to remain non-relativistic so that the radiation occurs in the cyclotron rather than the synchrotron regime.  The required pair multiplicity is $M_+ = 10^6 - 10^7$, well above what can be produced in either PC or OG pair cascades.  In our model, the pairs gain pitch angles through resonant absorption and acquire relativistic perpendicular momenta, so that the radiation is in the synchrotron regime.  Since the synchrotron radiation power is much higher than cyclotron power, we can produce the Crab optical to X-ray emission with a much lower pair multiplicity of $3 \times 10^5$.

%%%%%%%%%%%%%%%%%%%%%%%%%%%%%%%%%%%%%%%%%%%%%%%%%%%%%%%%%%%%%%%
\acknowledgments  %%%%%%%%%%%%%%%%%%%%%%%%%%%%%%%%%%%%%%%%%%%%%
%%%%%%%%%%%%%%%%%%%%%%%%%%%%%%%%%%%%%%%%%%%%%%%%%%%%%%%%%%%%%%%
We would like to thank Isabelle Grenier, Andrey Timokhin and Christo Venter for helpful discussions and acknowledge support from the NASA Astrophysics Theory Program and the $Fermi$ Guest Investigator Program.  Resources supporting this work were provided by the NASA High-End Computing (HEC) Program through the NASA Center for Climate Simulation (NCCS) at Goddard Space Flight Center.   We thank Craig Pelissier of the NCCS in particular for help with parallel processing.

\newpage

\newpage
 
\newpage

\newpage

\newpage

\newpage

\newpage

\newpage

%%%%%%%%%%%%%%%%%%%%%%%%%%%%%%%%%%%%%%%%%%%%%%%%%%%%%%%%%%%%%%%

\end{document}